\documentclass[10pt]{article}
\usepackage[a4paper, total={7in, 9in}]{geometry}
\usepackage[utf8]{inputenc}
\usepackage{graphicx}
\usepackage[dvipsnames]{xcolor}
\usepackage{amsmath,amssymb}
\usepackage{soul}
\usepackage{multirow}
\usepackage{authblk}
\usepackage{hyperref}
\usepackage{mathptmx}

\newcommand{\link}[1]{\href{#1}{#1}}

\begin{document}

\title{A statistical model to identify excess mortality in Italy in the period 2011-2022}
\author[1]{Antonino Bella}
\author[2,3]{Gianluca Bonifazi}
\author[4,5,6]{Luca Lista}
\author[7]{Dario Menasce}
\author[8]{Mauro Mezzetto}
\author[7]{Daniele Pedrini}
\author[1]{Patrizio Pezzotti}
\author[1]{Maria Cristina Rota}
\author[3,*]{Roberto Spighi}
\author[9,3]{Antonio Zoccoli}
\affil[1]{\raggedright\normalsize Istituto Superiore di Sanit\`a}
\affil[2]{\raggedright\normalsize Universit\`a Politecnica delle Marche}
\affil[3]{\raggedright\normalsize INFN Sezione di Bologna}
\affil[4]{\raggedright\normalsize Universit\`a degli Studi di Napoli Federico II}
\affil[5]{\raggedright\normalsize Scuola Superiore Meridionale, Naples, Italy}
\affil[6]{\raggedright\normalsize INFN Sezione di Napoli}
\affil[7]{\raggedright\normalsize INFN Sezione di Milano Bicocca}
\affil[8]{\raggedright\normalsize INFN Sezione di Padova}
\affil[9]{\raggedright\normalsize Alma Mater Studiorum Universit\`a di Bologna}
\affil[*]{\raggedright\normalsize \textbf{Corresponding author}, e-mail: {\tt roberto.spighi@bo.infn.it}}

\date{}

\maketitle

\begin{abstract}
Excess mortality is defined as an increase in the number of deaths above what is expected based on historical trends, hereafter called baseline.
In a previous paper, we introduced a statistical method that allows an unbiased and robust determination of the baseline to be used for the computation of excesses.
A good determination of the baseline allows us to efficiently evaluate the excess of casualties that occurred in Italy in the last 12 years and in particular in the last 3 years due to the Coronavirus Disease 2019 (COVID-19) epidemic.
To this extent, we have analyzed the data on mortality in Italy in the period January $1^{\mathrm{st}}$ 2011 to December $31^{\mathrm{th}}$ 2022, provided by the Italian National Institute of Statistics (ISTAT). The dataset contains information on deaths for all possible causes, without specific reference to any particular one.

The data exhibit strikingly evident periodicity in the number of deaths with pronounced maxima in the winter and minima in the summer, repeating itself in amplitude along the whole twelve-year sample. Superimposed on this wave-like structure are often present excesses of casualties, most likely due to occasional causes of death such as the flu epidemics (in winter) and heat waves (in summer). The very accurate periodicity along the seasons (the "baseline"), allows us to determine with great accuracy and confidence the number of expected deaths for each day of the year in the absence of occasional contributions. Each of the latter can be modeled with an additional function that parameterizes the deviation from the baseline. 
We can then provide an unbiased way of measuring these excesses for each period of the year and for different categories of people (disentangling the sample by age and gender). 
The study has been extended to include an analysis of the data provided by the Italian Department for Civil Protection (DPC) and the Italian National Institute of Health (ISS), which specifically focuses on deaths labeled as resulting from the COVID-19 pandemic.
The effect of COVID-19 on mortality is clearly visible since 2020; in Italy, the excess of deaths (until December 2022), directly or indirectly due to the COVID-19 pandemic, is estimated to be about 217 thousand distributed in 10 different pandemic waves. 
The majority of deaths (approximately 95\%) occurred in individuals over 60 years of age, and there is evidence indicating lower mortality rates among females. The excess of deaths has also been quantified by analyzing the data based on year, season, age group, and gender.

\end{abstract}

\section{Introduction}
The COVID-19 pandemic has emerged as one of the most devastating global health crises in recent decades, with a reported death toll of around 6.7 million worldwide (updated until December 2022). The advent of this pandemic has had a profound impact on people's daily routines and mindset. In this context, it becomes crucial to identify effective strategies to halt or at least reduce the spread of the epidemic without requiring drastic changes in human behavior. Equally important is the need to accurately assess the severity of the pandemic through reliable analytical methods. This would enable us to anticipate its development well in advance, providing policymakers with timely and well-informed data to undertake appropriate measures for mitigating its impact on the population.

The most accurate method for evaluating excess mortality is by determining the daily average expected death trend, which can be achieved by using a suitable model to interpolate data spanning a significant period. This data often exhibits notable periodicity. The key concept lies in the ability to assess the expected death rate caused by factors unrelated to the COVID-19 epidemic. This trend typically displays a wave-like behavior that can be mathematically extrapolated from the overall distribution with high accuracy. By employing a periodic function fit, it becomes possible to determine the expected death rate daily with great precision. Any deviation above this baseline curve can be attributed to seasonal effects such as COVID-19 (among other factors).

The excess mortality resulting from the COVID-19 pandemic encompasses individuals who have died directly from the virus, as well as those who were unable to receive treatment due to overwhelmed healthcare systems (particularly during the initial wave of the pandemic). A death is classified as COVID-19-related by the World Health Organization (WHO) when it is clinically compatible with a confirmed case of COVID-19 unless there is a clear alternative cause of death unrelated to COVID-19\cite{WHO_death}.
The analysis in this study relies on a least square fit approach to the data using a function consisting of a periodic component to account for the expected average number of deaths per day (the baseline), as well as a set of additional contributions. Each of these contributions is described by the derivative of the Gompertz function, which we will refer to as the
\emph{Gompertz function} throughout this paper.

This study is based on three different sets of data provided by ISTAT (Italian National Institute of Statistics)\cite{ISTAT-Data}, DPC (Italian Department for Civil Protection) \cite{ProtezioneCivile,ProtezioneCivile-Data} and ISS (Italian National Institute of Health)\cite{ISS-data}\cite{covidstat}. 

A first analysis has already been presented in \cite{dario} for a shorter period (January $1^{\mathrm{st}}$ 2015 to December $31^{\mathrm{th}}$ 2020), which contained only the first pandemic wave of the COVID-19 virus. 

The data have been carefully disentangled by gender and age classes, allowing for meaningful comparisons.

\section{The data sample}
\label{sec:data-sample}
The data provided by ISTAT consists of a time series of deaths for any cause recorded by the National Registry Office. The data, collected from all the 7904 districts located in the 20 Italian regions, cover the period from January $1^{\mathrm{st}}$ 2011 to December $31^{\mathrm{th}}$ 2022. The information collected comprehends the gender and the age of the dead people allowing to obtain a more detailed result. 

\begin{figure}[h]
    \centering
   \includegraphics[width=1.1\textwidth]{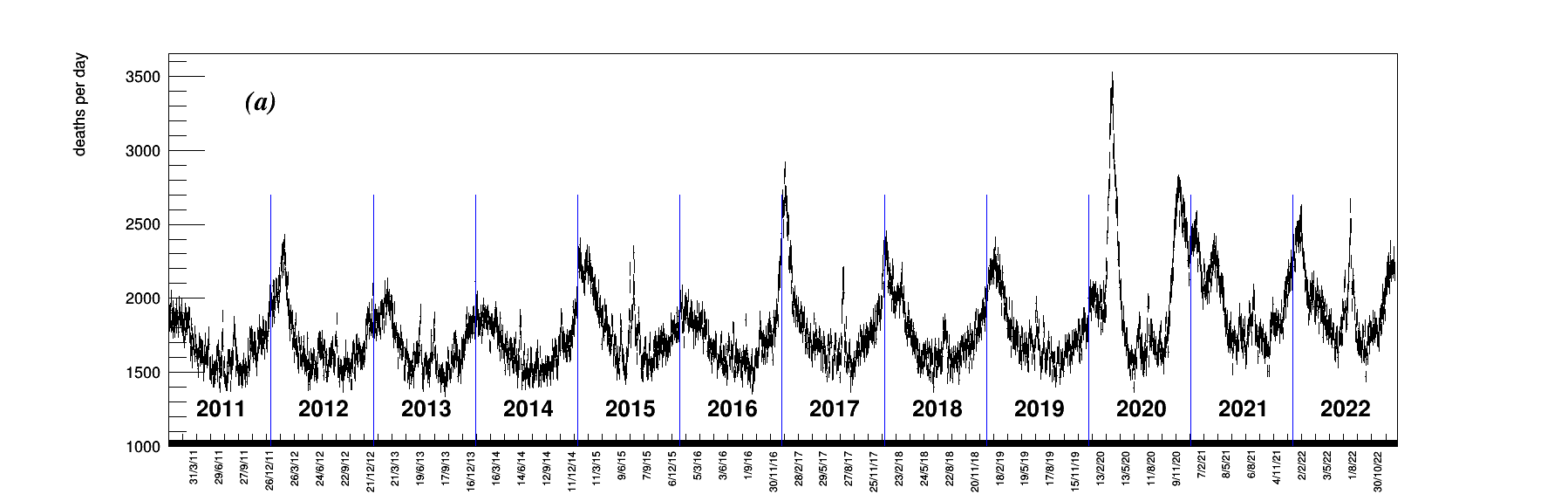}
   \includegraphics[width=1.1\textwidth]{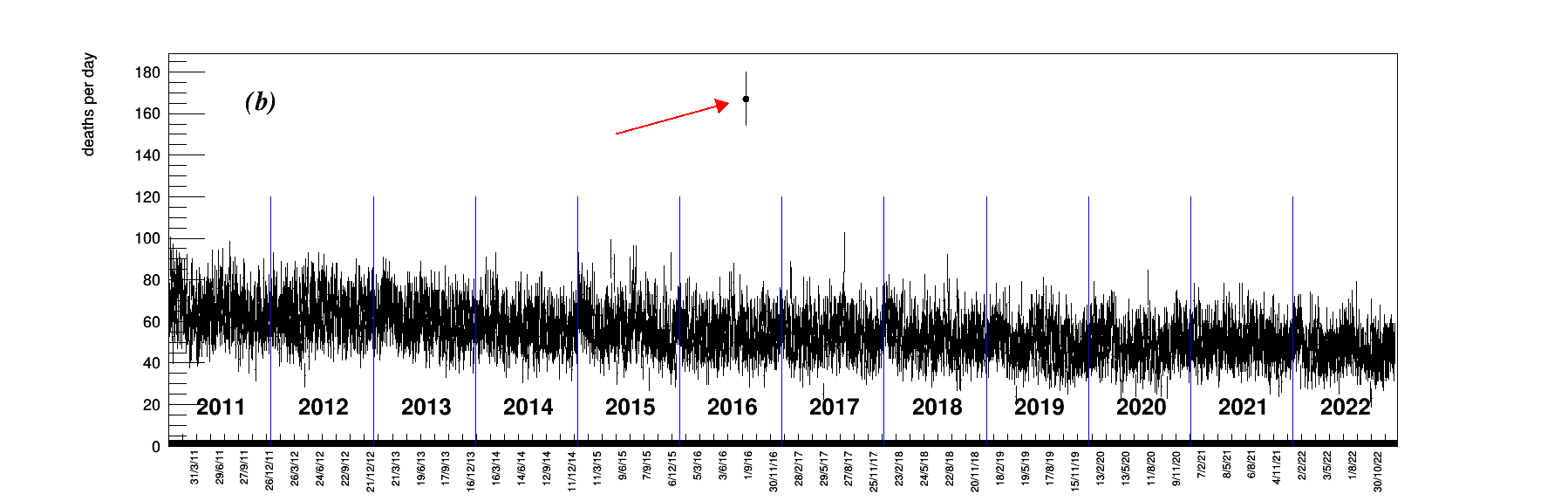}
    \caption{Distribution of any-cause-deaths collected by ISTAT \cite{ISTAT-Data} from 7904 districts in Italy between January $1^{\mathrm{st}}$ 2011 to December $31^{\mathrm{th}}$ 2022 for all classes of age (top) and the  0-49 class age only(bottom). In the bottom plot, marked by a red arrow) is clearly visible the tragic effect of the earthquake in Amatrice on August $24^{\mathrm{th}}$ 2016.}
    \label{fig:totalnofit}
\end{figure}

The crucial characteristics of the data from ISTAT are their independence from any specific cause of death and the accuracy of the recorded date of death. The data display a distinct wave-like pattern in the number of deaths, with prominent peaks occurring in the winter and troughs in the summer, as illustrated in Fig.~\ref{fig:totalnofit} (top).

The data obtained from the Italian Department for Civil Protection (DPC) and the Italian National Institute of Health (ISS) are specific to deaths directly attributed to COVID-19. The DPC data is provided daily, while the ISS data is derived from periodic reports. The DPC data, being available daily, reports the date of notification of death which may not coincide with the actual date of death and therefore may be subject to systematic effects such as inaccuracies in determining the date of death or reduced efficiency in data collection during weekends: these effects can negatively impact the quality of the fit. On the other hand, the ISS data, although not available daily, report the date of death and include a posteriori corrections of data.

It's important to note that both the DPC and ISS data use the same criteria to determine whether a death is attributed to the COVID-19 pandemic. These criteria may vary and introduce biases in the data.

Indeed, the data from the Italian National Institute of Health (ISS) is valuable as it provides information regarding gender and age class. Before conducting a statistical analysis, it is worth noting some observations that are already evident through a visual examination.
In Fig.~\ref{fig:totalnofit} (top), we observe a remarkable wave-like seasonal trend, with relative maxima every winter and minima in the summers. Applying a filter to the data, requiring an age between 0 and 49 
Fig.~\ref{fig:totalnofit} (down) the wave-like trend appears flattened out, indicating that maxima and minima are correlated with age. 

This pattern has been consistent over the past 12 years.

\section{Analysis method}
 \label{par:ana_met} 
We have performed a least-square fit of the data based on the $\chi^{2}$ minimization method, estimating simultaneously the baseline of the distribution and the excesses of deaths.
This method significantly differs from others reported in the literature~ \cite{OurWorldInData, HumanMortalityDatabase, WeeklyDeathStatistics,  temporalDynamics, Mannucci, Blangiardo,  crossRegionalAnalysis} where the baseline is subtracted by computing the average number of counts in the same period of the past 5 years. 
Expressing the excess of deaths relative to the average counts of previous years, only provides relative measurements and not absolute ones; when applying this method to years following a period of abnormally high excess deaths, it can result in a serious underestimation of the true and correct value.

 Other papers fit the baseline over the whole data including the excesses, either with just a simple straight function \cite{Karlinsky} or with a straight function plus harmonics \cite{Henry} or by using external packages \cite{outbreakMonitorig}. While these latter methods can capture the periodicity and the phase of the baseline, they have severe limitations in reproducing its amplitude, which is hidden behind the seasonal excesses by flu (winter) or heat waves (summer).

A slightly different method, with similar limitations regarding the baseline amplitude, is used in \cite{Euromomo}, where the baseline is computed by using the periods where the seasonal effects are not visible.

 Bayesian hierarchical models are also used to estimate the baseline \cite{excessUS, Woolf, Kontis}. These methods partially attenuate the influence of epidemics in the historical record, at the price of quite large statistical errors.

A global fit to the time series has, instead, the merit of simultaneously using all the available data to shape both the excesses and the baseline, provided a meaningful model adopted to describe the data analytically. 
Furthermore, the least-squares method provides a complete and fully correct covariance matrix that allows for the computation of the uncertainties of the parameters involved in the final result. The goodness of the fit and the absence of biases are quantified by the final $\chi^2$ value of the interpolation and the pull distribution respectively (see paragraph~\ref{par:fit_quality} for their definition). 
The actual $\chi^{2}$ minimization is carried out by the MINUIT~\cite{MINUIT} package, while the adopted statistical methodology is described in~\cite{James}. The data are assumed to follow a Poisson distribution, hence the uncertainty associated with each daily number is the square root of the number itself. All the data have been fitted also with the maximum likelihood estimation, obtaining fully compatible results that will not be reported so as not to burden the article.
 The analytical form of the function chosen to model the distribution of death is arbitrary and this choice could introduce a systematic error in the determination of the number of deaths. For this reason, we fit the data with 6 slightly different fit functions to estimate the systematic contribution arising from the specific choice of the aforementioned function. In general, is important to use a function with the smallest number of parameters as allowed by the theoretical model: it is always possible to obtain a better fit of the data sample with an arbitrarily large number of free parameters but this comes at the cost of introducing a decreasing adherence to a meaningful theoretical model.

The fit function $F_{ref}(t)$ (where "ref" stands for reference function) has been factorized into two separate components: the first represents the baseline of expected deaths for each day, assuming no particular cause is involved, while the second is a sum of functions designed to appropriately represent excesses from the norm. By separating these two components, the fit function can capture both the expected baseline and the deviations from it.

\begin{equation}
  \label{fullFitFunction}
  F_{ref}(t)=baseline(t)+
  {\sum_{j=1}^{n}\dot{G}_{j}(t)}
\end{equation}

where $t$ is the day number from  January $1^{st}$ 2011 to December $31^{th}$ 2022 and $\dot{G}_{j}(t)$ are the Gompertz functions \cite{Gomp} used to model all the "$n$" excesses of deaths (see below for its definition).
The $baseline(t)$ function represents the number of expected deaths for each day if no external cause occurs to create an excess concerning the average trend.
It is expressed by the sum of a straight line, to describe population aging over a twelve-year long period and a periodical contribution for the wave-like variation of deaths along the passing seasons, in the form: 
\begin{equation}
  \label{baseline}
  baseline(t)=const + slope\cdot t + (A_0+A_1\cdot t) \sin\left(\frac{2\pi t}{T}+\varphi\right)\,
\end{equation}

The free parameters of the baseline are:
\begin{itemize}
\item $const$ which represents the average number of deaths per day;
\item $slope$ which describes the increase in the elderly population;
\item $A_0$ and $A_1$, which represent the amplitude of the sinusoidal, with $A_1$ taking into account the increase in the population subject to seasonal variation;
\item $T$ is the period, the time between two consecutive maxima or minima;
\item $\varphi$ is the phase of the oscillation.
\end{itemize}
It is clear that the parameters $const$, $slope$, $A0$ and $A1$ are strongly correlated, and alternative functional forms could equally describe the data well. The choice fell on this combination because it has the best $\chi^{2}/{n_{\mathrm{DOF}}}$ value.
In the $F_{ref}(t)$ reference fit, each excess of death has been modeled by a Gompertz function, a sigmoid function commonly used to describe growth processes (not only pandemic but also cancer, population, fetus development, ...); 
we have used its derivative to fit the excess of daily counts of the death distribution and it is defined  as:
\begin{equation}
    \dot{G}(t;a,k_G,T_p) = C \cdot k_{\tiny G} e^{-e^{-k_{\tiny G}(t-T_p)}-k_{\tiny{G}}(t-T_p)}\,\,\,.
    \label{eq:gomp}
\end{equation}
where in this parametrization \cite{Gomp}, $C$ is the integral of the curve (the number of deaths under the curve), $T_p$ the time when the peak is reached and $K_G$ the growth-rate coefficient modeling the shape: the inverse of $K_G$ multiplied by 2 is a good approximation of the full width at half maximum. $C$, $T_p$, and $K_G$ are the free parameters of the Gompertz function.
An advantage of this function is the small number of free parameters and in the case of adjacent overlapping of Gompertz functions, each area is computed correctly by considering the nearby contributing ones. 

To identify the excess death peaks, for each day $i$ it has been defined a Significance function $S(i)$ as
\begin{equation}
    S(i) = \frac{N_{deaths}(i)-N_{baseline}(i)} {\sqrt{N_{baseline}(i)}}
    \label{eq:signif}
\end{equation}
where $N_{deaths}(i)$ is the effective number of deaths, $N_{baseline}(i)$ is the expected ones in the absence of any external cause (flu, COVID-19, heat wave, ...) and the denominator represent its poissonian fluctuation. An excess of death has been defined as if on at least 3 consecutive days they have $S(i) > 4$. In the overall period, 60 significant excess death peaks have been identified thus requiring a total number of 186 free parameters: the number of degrees of freedom ( $n_{DOF}$) is 4195 indicating a very constrained fit.

The data reported by the DPC and ISS are relative to the deaths directly attributed to COVID-19, so the distributions have been modeled only with a sum of the Gompertz function without the baseline. In both data, several peaks of deaths are visible, which have been used to identify the position of the corresponding excesses due to COVID-19 in the data from ISTAT.

\section{Results and discussion}
\begin{figure}[h]
    \centering
    \includegraphics[width=1.\textwidth]{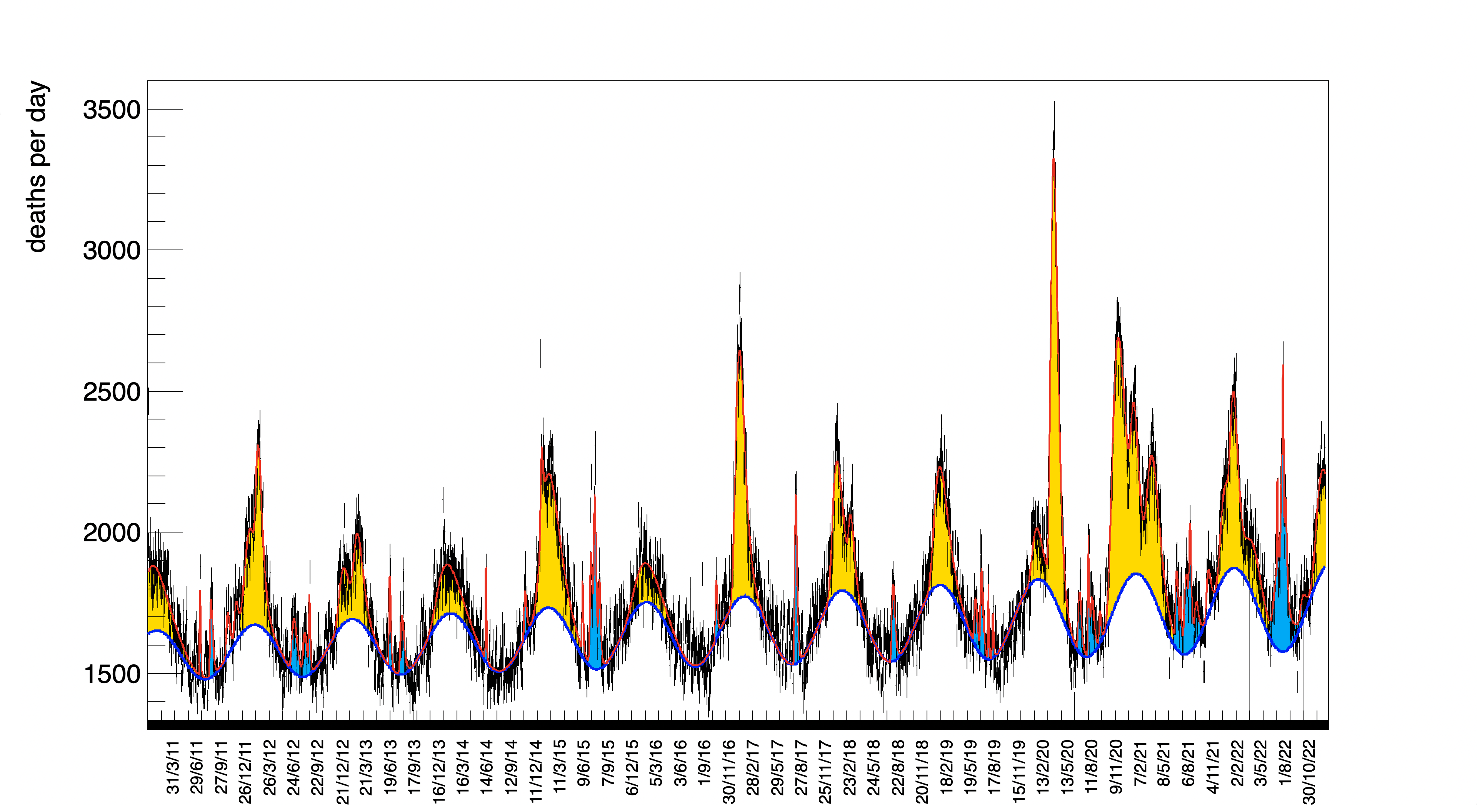}
    \caption{Distribution of any-cause-deaths collected by ISTAT \cite{ISTAT-Data} from 7904 districts in Italy between January $1^{\mathrm{st}}$ 2011 to December $31^{\mathrm{th}}$ 2022. These data include both genders and all classes of age. The distribution is well modeled by the fit function~\ref{fullFitFunction} (red curve), with superimposed the baseline (blue curve).
    The yellow area signifies the excess mortality compared to the baseline curve in winter, whereas the cyan area denotes the excess mortality observed in summer.
    }
    \label{fig:totalfit}
\end{figure}

Figure~\ref{fig:totalfit} shows the distribution of the ISTAT data sample relative to all deaths from any cause that occurred in Italy from January $1^{\mathrm{st}}$ 2011 to December $31^{\mathrm{th}}$ 2022, fitted with the reference function (eq.~\ref{fullFitFunction}). 

The distribution shows a periodic trend corresponding to the baseline (blue line of the fit function) from which to evaluate all the excesses of death: the parameters of the baseline obtained after the convergence procedure of the fit are reported in Table~\ref{table:baseline}.

 \begin{table}[htbp]
 \centering
 \begin{tabular}{|| c | c | c | c | c | c ||} 
  \hline
  \emph{$const$} & \emph{$slope$} &\emph{$A_0$} & \emph{$A_1$} &\emph{$T$} (days) & \emph{$\varphi$} (rad)\\ [0.5ex] 
  \hline\hline
 1561.8 $\pm$ 3.9 & 0.040 $\pm$ 0.002 & 89 $\pm$ 5 & 0.015 $\pm$ 0.002 & 364.7 $\pm$ 0.2 & -5.23 $\pm$ 0.02 \\ \hline
 \end{tabular}
 \caption{Output parameters of the baseline obtained fitting the whole data set (no selection applied), as modeled by eq.~\ref{fullFitFunction}. }
 \label{table:baseline}
 \end{table}

It is worthwhile to emphasize a few key aspects:
\begin{itemize}
\item in Italy about 1,600 people died, on average, every day in the last 12 years;
\item the number of deaths per day has slightly increased (a fact represented by a $slope >0$). On average, about 100 more people died every day in 2022 than in 2011: an explanation could be the increase of the average age of the population;
\item on average, about 270 more people die every day in winter than in summer;
\item the value of the period $T$ is remarkably compatible with a full year cycle;
\item the amplitude of the sine function increases over time;
\item the maximum number of expected deaths (the peak of the baseline) falls about on January $31^{st}$ of each year. 
\end{itemize}

\begin{figure}[htbp]
    \includegraphics[width=0.5\textwidth]{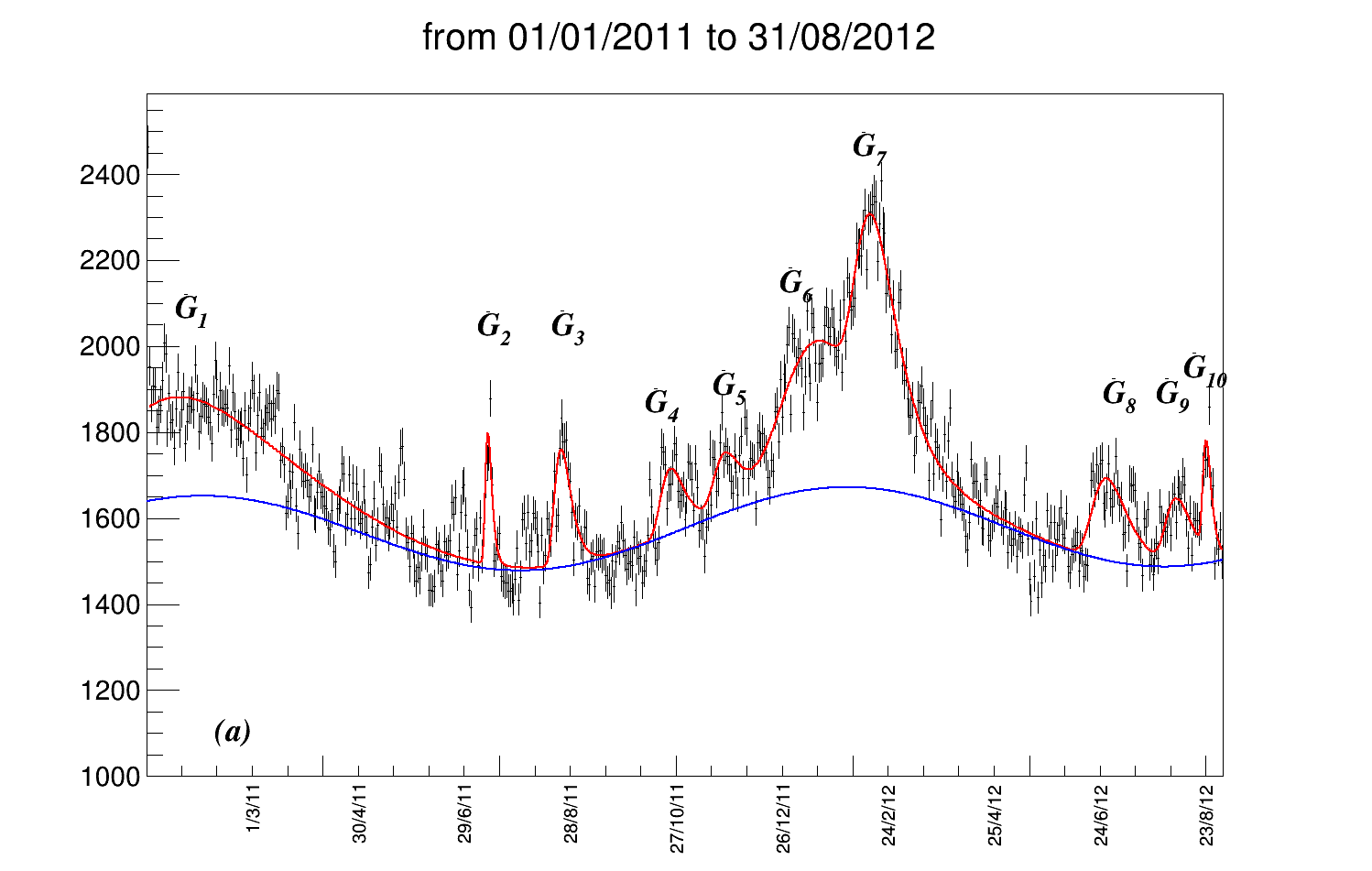}
   \includegraphics[width=0.5\textwidth]{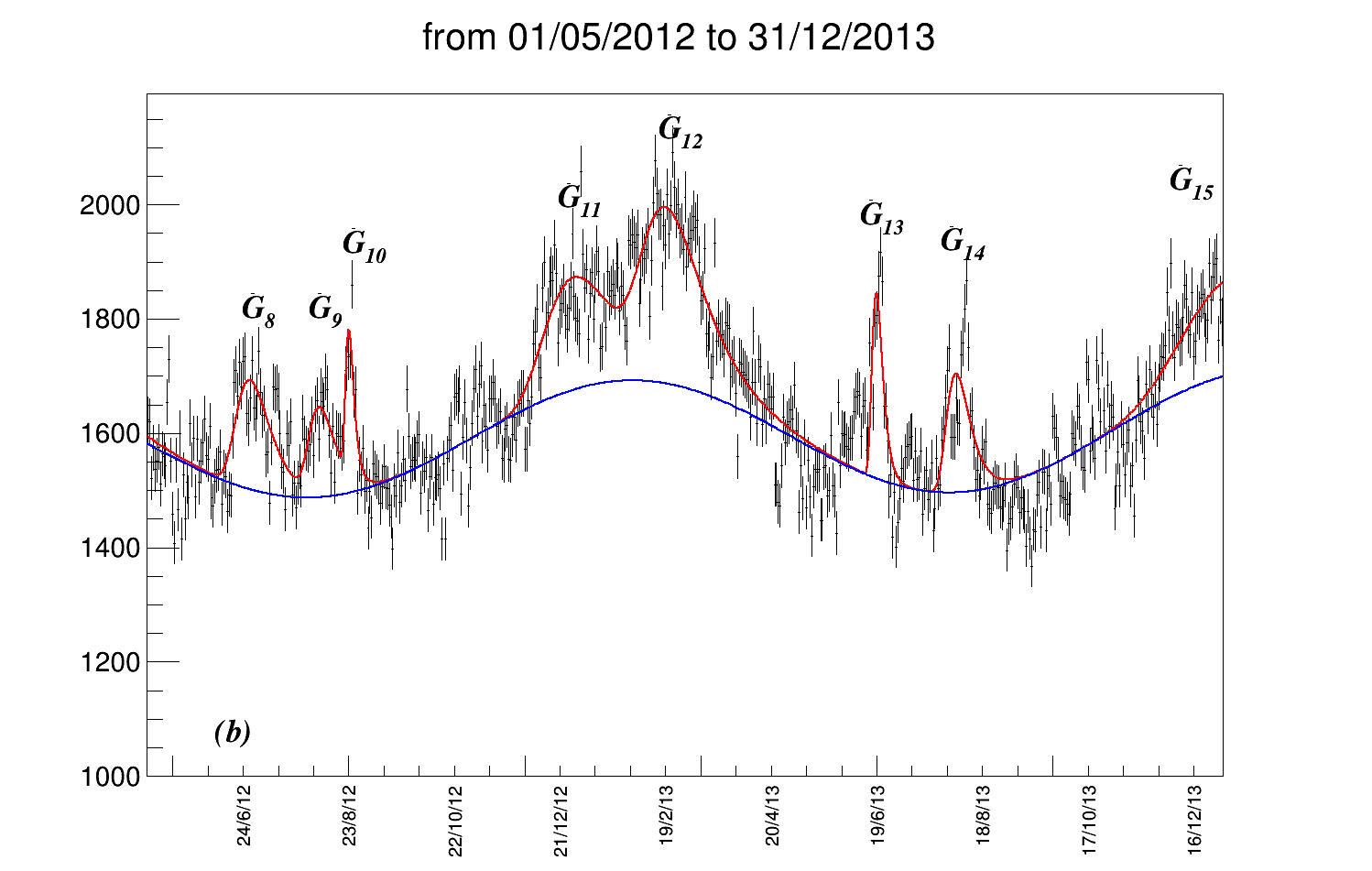}    
    \includegraphics[width=0.5\textwidth]{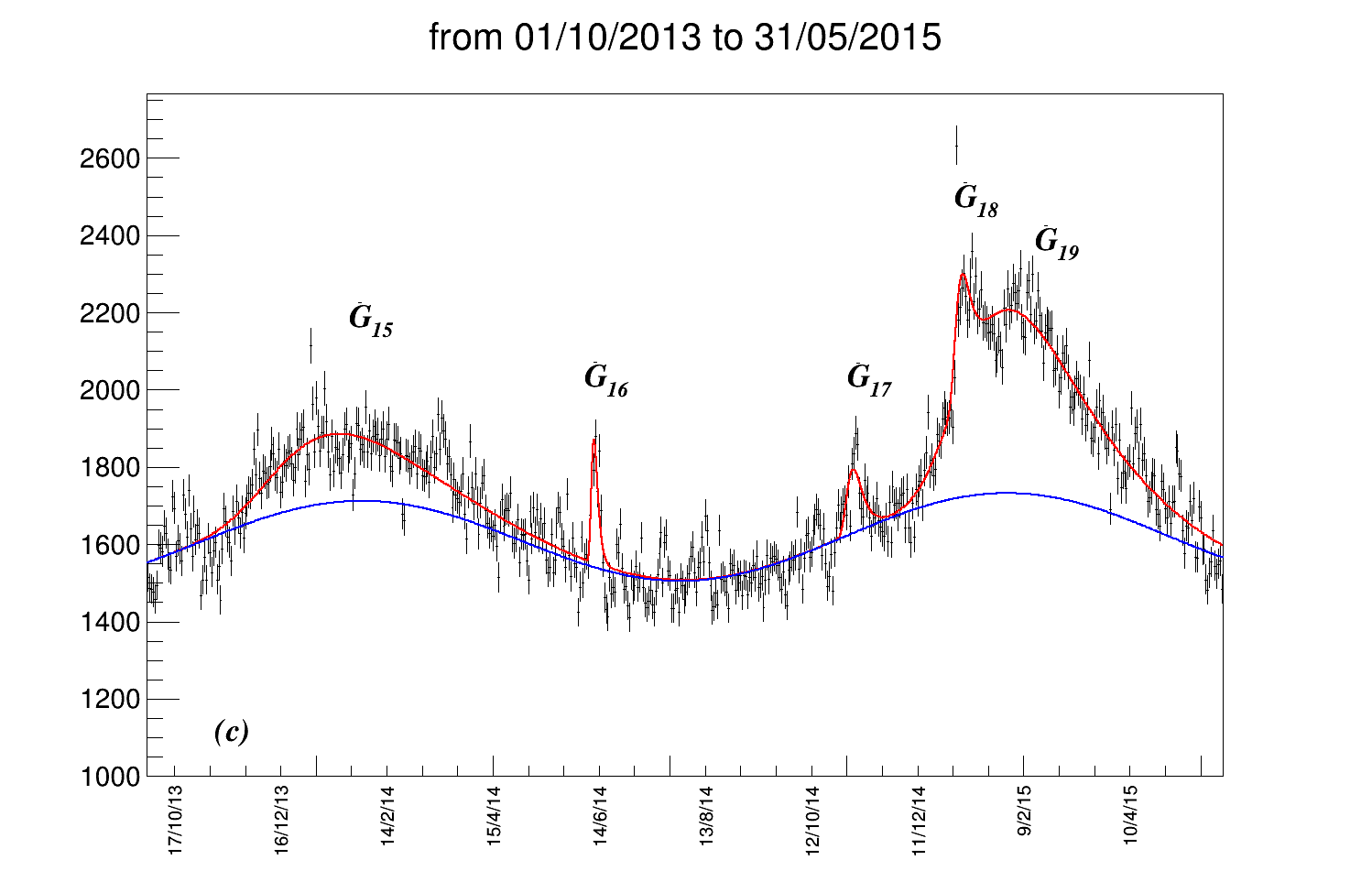}
    \includegraphics[width=0.5\textwidth]{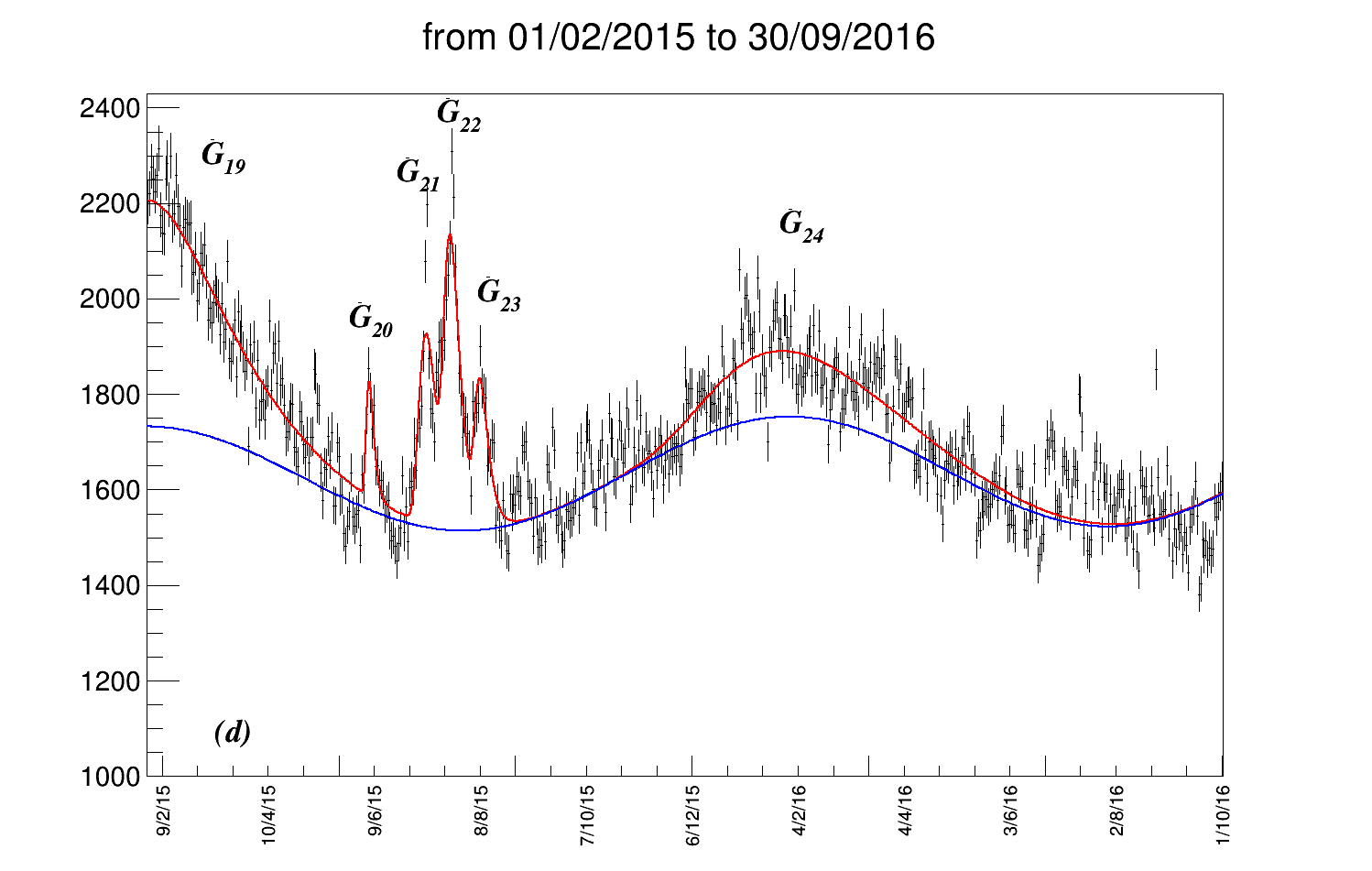}
    \includegraphics[width=0.5\textwidth]{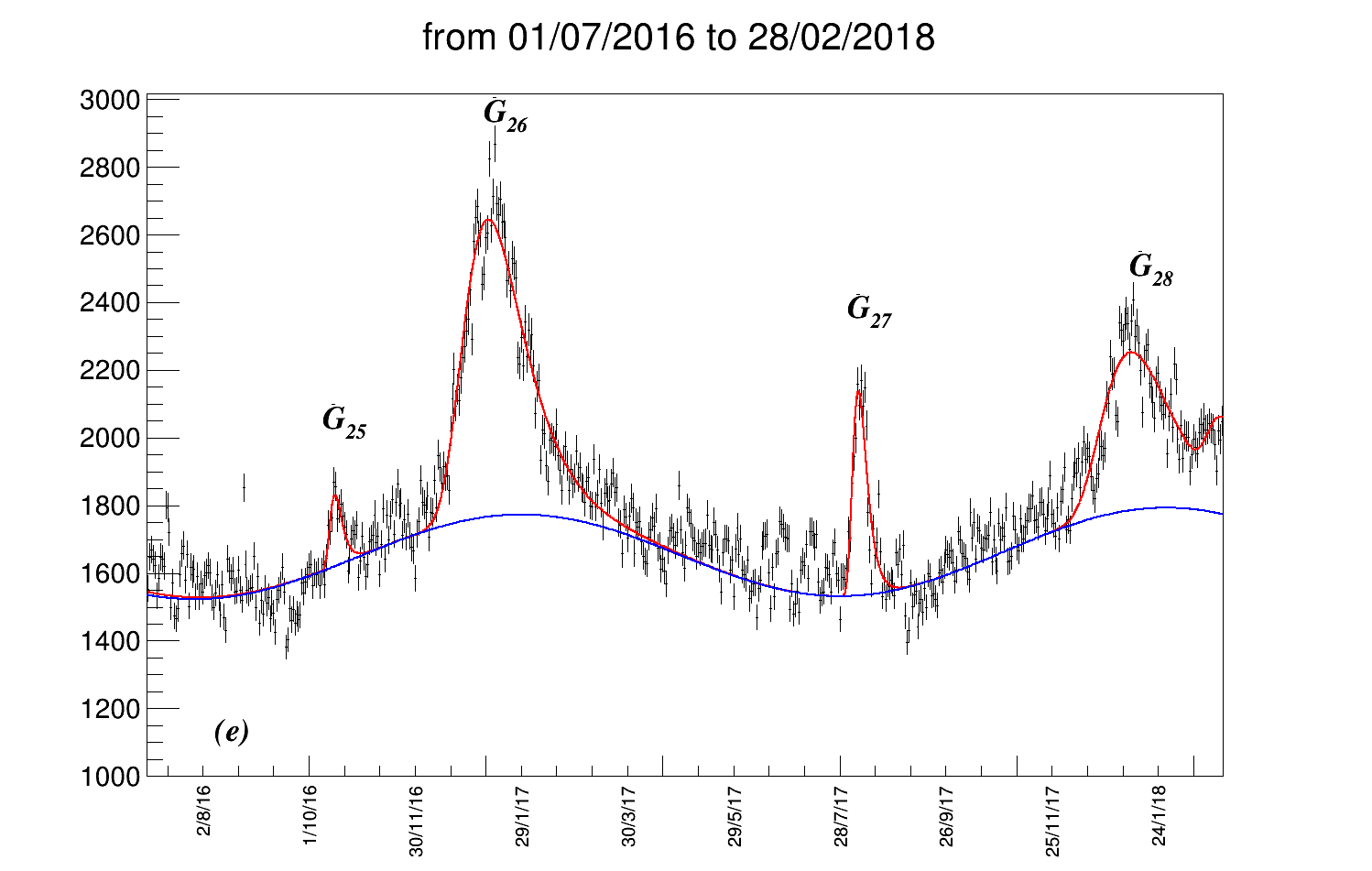}
    \includegraphics[width=0.5\textwidth]{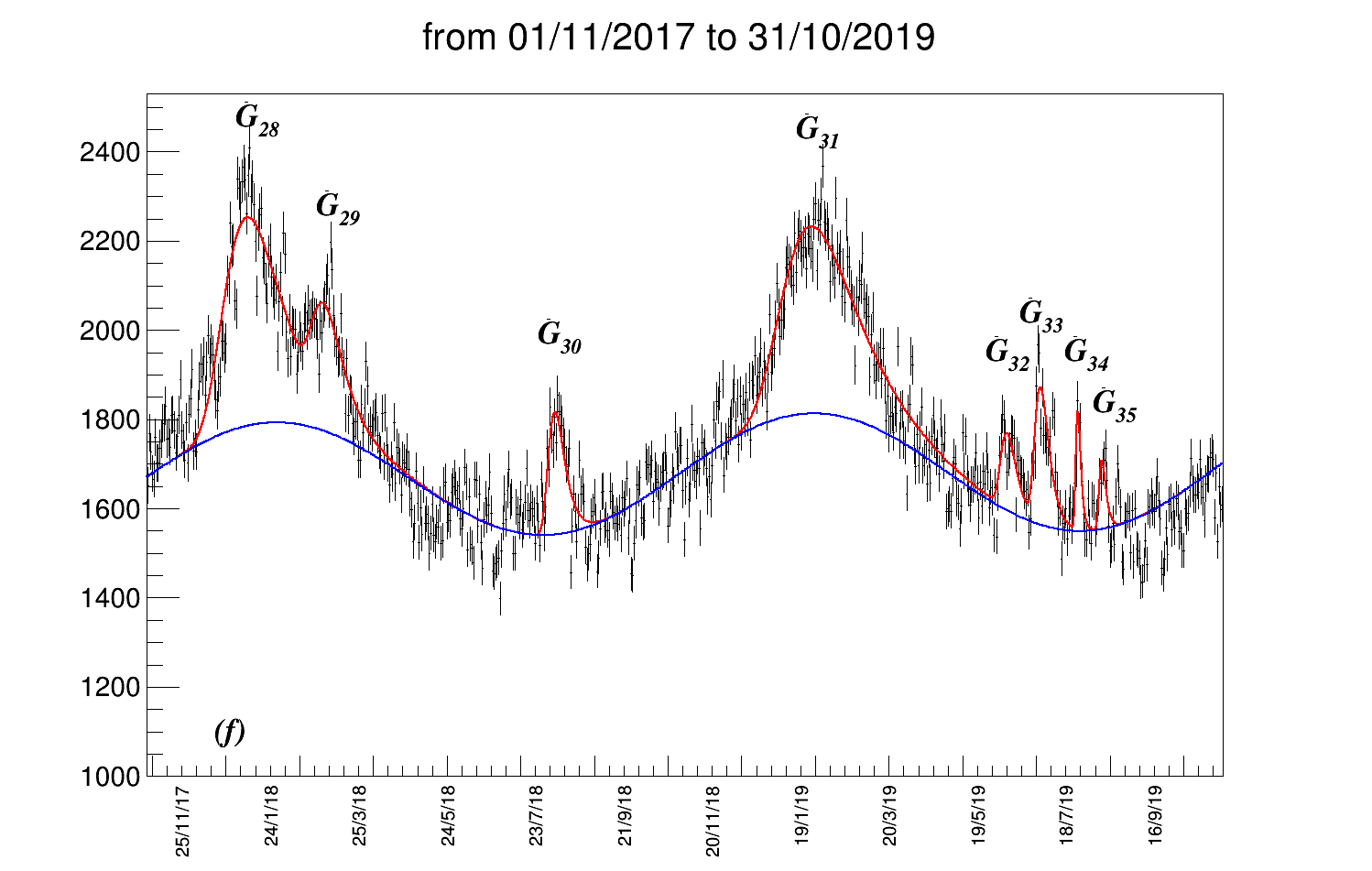}
    \caption{Distribution of deaths divided into 6 subsequent \st{time} periods between January $1^{\mathrm{st}}$ 2011 to October $31^{\mathrm{st}}$ 2019. Each Gompertz function is indicated with $\dot{G}$. To enhance the readability of these plots, we produced them with a significant time overlap between adjacent ones.}
    \label{fig:zoomfit}
\end{figure}

For better visualization of the data and of the interpolated function, all the periods, previous to the COVID-19 pandemic, have been divided into 6 subsequent time intervals (with a significance time overlap) and shown in Figure~\ref{fig:zoomfit}. 

The excesses of death in the period autumn-winter-spring and in summer have been summarized in  Table~\ref{tab:wint_sum}.
 \begin{table}[htbp]
 \centering 
 \begin{tabular}{|| c | c | c | c ||} 
  \hline
   Autumn-Winter-Spring & death excess $\pm$ stat $\pm$ syst & Summer & death excess \\ [0.5ex]
  \hline\hline 
    2010-11  &  $27461 \pm  2224 \pm  2952 $  &   2011  & $4784 \pm  274 \pm  201 $ \\ 
    2011-12  &  $40828 \pm  2203 \pm  5273 $  &   2012  & $9282 \pm  409 \pm  439 $ \\ 
    2012-13  &  $22795 \pm  857 \pm  1908 $  &   2013  & $6235 \pm  265 \pm  579 $ \\ 
    2013-14  &  $18689 \pm  823 \pm  1716 $  &   2014  & $1540 \pm  114 \pm  32 $ \\ 
    2014-15  &  $46916 \pm  912 \pm  2069 $  &   2015  & $15861 \pm  679 \pm  1119 $ \\ 
    2015-16  &  $16428 \pm  880 \pm  1085 $  &   2016  & no excesses of deaths \\ 
    2016-17  &  $41719 \pm  535 \pm  1100 $  &   2017  & $6452 \pm  190 \pm  178 $ \\ 
    2017-18  &  $29818 \pm  687 \pm  1542 $  &   2018  & $4072 \pm  191 \pm  112 $ \\ 
    2018-19  &  $27709 \pm  767 \pm  1564 $  &   2019  & $8302 \pm  313 \pm  289 $ \\ 
    2019-20  &  $62909 \pm  846 \pm  2022 $  &   2020  & $10061 \pm  353 \pm  1009 $ \\ 
    2020-21  &  $105469 \pm  1225 \pm  5553 $  &   2021  & $17123 \pm  510 \pm  1294 $ \\ 
    2021-22  &  $50894 \pm  2316 \pm  3790 $  &   2022  & $27776 \pm  1558 \pm  1781 $ \\ 
    2022*  &  $26655 \pm  1526 \pm  3338 $  &   2023  &  \\ 
  \hline\hline
 \end{tabular}
 \caption{Excesses of death (together with the statistical and systematic uncertainty) in summer and the rest of the year. (*)The excess of deaths has been computed only up to December $31^{\mathrm{th}}$ 2022.}
 \label{tab:wint_sum}
 \end{table}

Some points deserve additional comments here:
\begin{itemize}
\item in every season there is an excess of deaths compared to the baseline. 
It is unequivocal that after the outbreak of the COVID-19 pandemic, the excess of deaths in the winter period (when the impact of the virus is most severe) has significantly increased, passing from an average of about 30 thousand excess deaths from 2011 to 2019 to about 73 thousand between 2020 and 2022 (see Table~\ref{tab:wint_sum}). 
In addition to COVID-19, the causes of these excesses can stem from various reasons, such as colder temperatures or seasonal diseases like influenza. In the last 12 years, the effect of the temperature is clearly visible: the mildest winters have been 2012-13, 2013-14 (warmest of all), and 2015-16, confirmed by a lower excess of deaths. On the contrary, in February 2012, December 2014, and February 2018 there were extraordinary waves of frigid temperatures that led to an excess of deaths visible as steep peaks labeled G7, G18, and G28 in Figure~\ref{fig:zoomfit}-a-c-f respectively. 
\item in summer, the typical excesses of deaths are very steep and short-lived because they are the effect of heat waves, which usually last a few days. One of the strongest series of heat waves of the last 100 years (after the record of 2003 not present in these data) occurred in the summers of 2015, 2021, and 2022: the high number of deaths for those periods is reported in Table~\ref{tab:wint_sum} and, regarding summer 2015, is clearly visible as a series of pronounced peaks in Fig.~\ref{fig:zoomfit}-d (summers 2021 and 2022 are shown in the figure of the next paragraph).  
\end{itemize}

We will compare our results during two specific periods, with prominent peaks in excess mortality, with estimates obtained using the three methods previously employed in medical research:

\begin{enumerate}
\item the mean number of deaths calculated by averaging deaths recorded in the previous 5 years;
\item the mean number of deaths calculated by averaging the number of deaths for each day in the preceding 5 years;
\item a log-linear regression function. This function is the basis of the "Bayesian Hierarchical Poisson Regression Model for Overdispersed Count Data", which is a statistical model commonly used in medical research for analyzing count data that shows more variation than expected by the Poisson model.
\end{enumerate}
Table~\ref{tab:ave_meth} shows the results obtained with these three methods compared to our estimate.

 \begin{table}[htbp]
 \centering 
 \begin{tabular}{|| c | c | c ||} 
  \hline
  \emph{methods}  & \emph{Excess Dec 2016 Jan-Feb-Mar 2017} & \emph{Excess Mar-Apr-May 2020} \\
  \hline\hline
  1 method   &  $\sim 47100$   &   $\sim 49100$  \\
  2 method   &  $\sim 23000$   &   $\sim 50900$  \\
  3 method   &  $\sim 48400$   &   $\sim 57600$  \\
  our estimate &  $39830 \pm 512 \pm 1064$ &  $53615 \pm 508 \pm 809$ \\
  \hline\hline
 \end{tabular}
 \caption{Our estimations of the excess of deaths obtained with different methods present in literature.}
 \label{tab:ave_meth}
 \end{table}

\begin{figure}[h]
    \centering
    \includegraphics[width=0.7\textwidth]{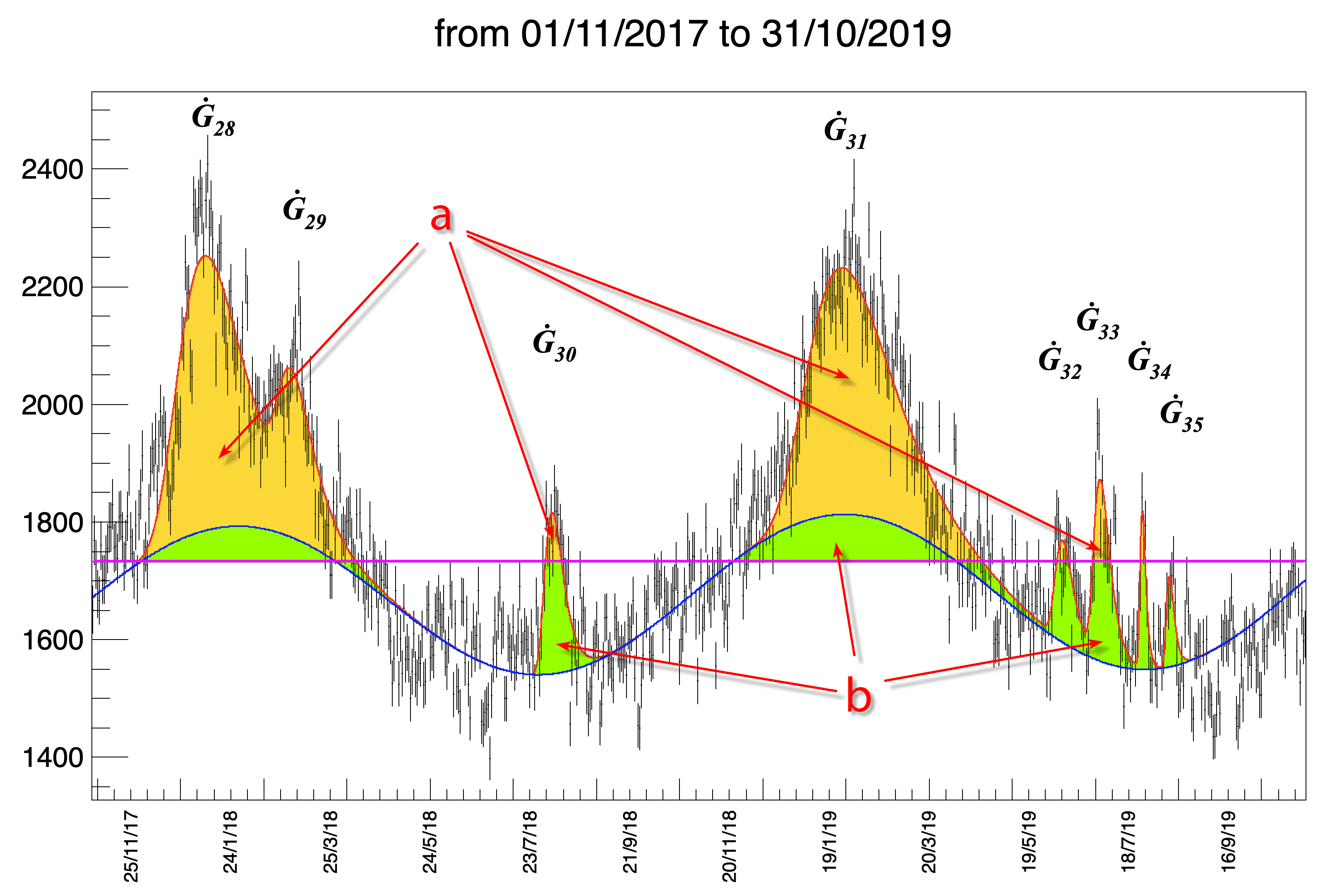}
    \caption{Rate of deaths in the years 2017/2019. The horizontal violet line is the value of the simple average of all the deaths that occurred in the preceding five years (2013/2017), while the blue line is the parametrization of the seasonal deaths with our baseline. Using the simple average as an estimate of the typical number of deaths, the excess would be computed as the sum of the yellow areas PLUS the green one during the winter periods (thus overestimating the excess, such as for $G_{28}$ and $G_{29}$). On the contrary, in the summer periods, the excess would be computed with an underestimate of the correct number, such as $G_{30}$, indicated by the cyan color.
    The label $a$ indicates the mortality excess evaluated by the fit function above the baseline, while the label $b$, on the contrary, indicates the overestimation component of the mortality excess in winter (and the underestimate in the summer) if the baseline is assumed to be the average of the preceding years.}
    \label{fig:averageLine}
\end{figure}

We have noticed a significant difference between our determination and the results obtained by the other three methods. The three methods, although reasonable in some contexts, are not suitable for this particular case, because the linear trend of the function underlying the three models cannot capture the sinusoidal shape, evident in the raw data.
Figure~\ref{fig:averageLine} shows how better a wave-like model for the baseline allows the extraction of the correct excess of deaths with respect to a plain average. The average number of deaths computed for the years 2013-2017 is shown as a horizontal line, amounting to 1733 deaths per day. The number of excess deaths computed using this value is shown in the picture as the yellow areas (marked by the letter $a$, corresponding to the number of excesses with respect to the sinusoidal parametrization of the baseline) plus the green one (the $b$ areas). It is self-evident that this number erroneously includes a significant number of deaths which should instead be subtracted from the total, since the seasonal effects, modeled by the blue line, clearly show a natural increase in deaths in winter and a decrease in the summer. 

\subsection{Zoom on the COVID-19 pandemic period}
 \label{par:covid_waves} 
\begin{figure}[h]
    \includegraphics[width=1\textwidth]{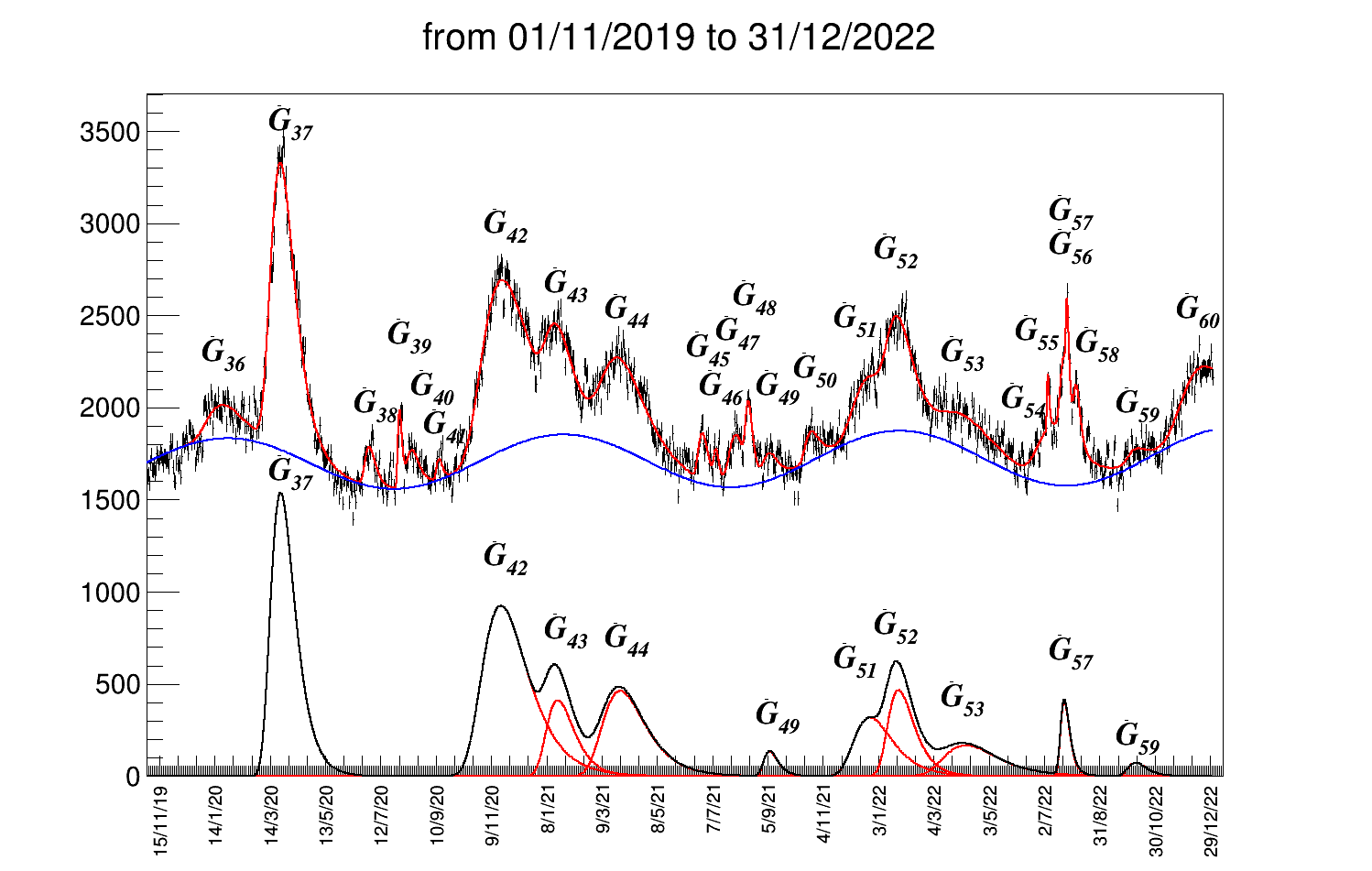}
    \caption{Distribution of deaths collected from November $1^{\mathrm{st}}$ 2019 to December $31^{\mathrm{th}}$ 2022: in the top part all the functions are indicated, in the bottom part only the ones due to the COVID-19 pandemics (in red the single waves and in black the total, where the baseline has already been subtracted).}
    \label{fig:covidfit}
\end{figure}

Figure~\ref{fig:covidfit} shows the distribution of deaths from November 2019 to December $31^{\mathrm{th}}$ 2022, to better visualize the COVID-19 period.

In Italy, the outbreak of the COVID-19 pandemic was in February 2020 and the consequences of the excess deaths are visible in Fig.~\ref{fig:covidfit} and in Table~\ref{tab:death_year}, 
where are reported the total number of deaths per year: in 2020, 2021 and 2022 the number of deaths exceeds by about 20\% the average of the previous years.

It is important to establish an unbiased strategy to evaluate how many of these deaths are directly or indirectly caused by the COVID-19 pandemic and to this extent, we have identified the peaks due to COVID-19 fitting the data from DPC and ISS (see paragraph  \ref{par:dpc_iss}), that refers only to death attributed to COVID-19 disease. 
To minimize the possibility of wrong identification, we have defined an excess of deaths due to COVID-19 if and only if there is a peak in the data provided by both the ISS and DPC in the same period, i.e. there is a temporal correlation. A similar temporal correlation has been adopted to evaluate the excess of deaths in US counties due to COVID-19 \cite{Paglino}.

 \begin{table}[htbp]
 \centering 
 \begin{tabular}{|| c | c | c | c | c | c | c | c | c | c | c | c ||} 
  \hline
   2011 & 2012 & 2013 & 2014 & 2015 & 2016 & 2017 & 2018 & 2019 & 2020 & 2021 & 2022\\ [0.5ex]
  \hline\hline 
    608575 & 624888 & 610608 & 608953 & 656196 & 627071 & 659473 & 640843 & 644515 & 746146 & 709035 & 713499\\
    \hline
 \end{tabular}
 \caption{Total number of deaths per year counted by ISTAT \cite{ISTAT-Data} }
 \label{tab:death_year}
 \end{table}

From February 2020 until December 2022, we have identified ten different excesses of deaths (in some cases partially overlapped) in the data provided by ISTAT due to the COVID-19 disease; an advantage of the fit approach is that
in the case of adjacent, overlapping functions, each area is computed correctly by taking into account the nearby contributing ones.  In Table~\ref{tab:covpar} all the main characteristics of the COVID-19 pandemic waves have been reported.

 \begin{table}[htbp]
 \centering 
 \begin{tabular}{|| c | c | c | c | c | c ||} 
  \hline
  \emph{Period}  & \emph{function} & \emph{$T_{p}$} (dates)  & \emph{ISTAT} & \emph{DPC} &\emph{ISS} \\ [0.5ex] 
  \hline\hline 
  March 2020     &   $\dot{G}_{37}$   &   24/3/2020   &   $53615 \pm 508  \pm 822$    &  $33451 \pm 183 $   &   $34769 \pm 185 $  \\
  November 2020  &   $\dot{G}_{42}$   &   18/11/2020  &   $60855 \pm 743  \pm 4506$   &  $48031 \pm 430 $   &   $51997 \pm 500 $  \\
  January 2021   &   $\dot{G}_{43}$   &   19/1/2021   &   $16348 \pm 591  \pm 2898$   &  $15732 \pm 524 $   &   $17214 \pm 684 $  \\
  March 2021     &   $\dot{G}_{44}$   &   28/3/2021   &   $28266 \pm 773  \pm 1462$   &  $24266 \pm 354 $   &   $22610 \pm 436 $  \\
  August 2021    &   $\dot{G}_{49}$   &   6/9/2021    &   $2700 \pm 227  \pm 739$     &  $4021 \pm 120 $    &   $3726 \pm 101 $   \\
  October 2021   &                    &   21/10/2021  &   $3176 \pm 237  \pm 237$     &  $2155 \pm 620 $    &   $1295 \pm 305 $   \\
  December 2021  &   $\dot{G}_{51}$   &   23/12/2021  &   $17569 \pm 1599  \pm 3067$  &  $11288 \pm 2026 $  &   $8993 \pm 442 $   \\
  January 2022   &   $\dot{G}_{52}$   &   24/1/2022   &   $17455 \pm 1385  \pm 1428$  &  $12455 \pm 1228 $  &   $16182 \pm 621 $  \\
  March 2022     &   $\dot{G}_{53}$   &   7/4/2022    &   $12694 \pm 913  \pm 1691$   &  $8354 \pm 462 $    &   $8644 \pm 210 $   \\
  July 2022      &                    &   25/7/2022   &   $1565 \pm 619  \pm 207$     &  $7474 \pm 134 $    &   $7419 \pm 221 $   \\
  July 2022      &   $\dot{G}_{57}$   &   22/7/2022   &   $5288 \pm 1038  \pm 640$    &  $328 \pm 34 $      &   $700 \pm 235 $    \\
  October 2022   &   $\dot{G}_{59}$   &   8/10/2022   &   $1886 \pm 297  \pm 213$     &  $3003 \pm 400 $    &   $3337 \pm 334 $   \\
  December 2022  &                    &   19/12/2022  &   $24769 \pm 1497  \pm 3331$  &  $6970 \pm 873 $    &   $6934 \pm 740 $   \\
  Total         &                     &               &   $216677 \pm 2878 \pm 6843$  &  $177528 \pm 2790 $ &   $183821 \pm 1560$ \\
  \hline
 \end{tabular}
 \caption{The number of deaths for each COVID-19 pandemic wave has been obtained from the fit parameters of the Gompertz functions on the whole data set (no filters applied). In the last two columns, the number of deaths obtained by the DPC and ISS data is reported respectively with only statistical uncertainty, while for ISTAT we also report the systematic error as the second error value.}
 \label{tab:covpar}
 \end{table}

The data provided by ISS and DPC have been interpolated with thirteen Gompertz functions: three of these were not identified in the ISTAT data as they contained a small number of deaths and as they were temporally close to various summer excesses due to heat waves. 

From the ISTAT data, until December 2022, the total number of excess deaths of all the COVID-19 pandemic waves, amounts to $ 216677 \pm 7424 $ (including the systematic error see Table~\ref{tab:covpar}); this number includes the people who died directly from COVID-19 and those who died from the indirect effects of the pandemic. In the same period, from the data provided by DPC and ISS, the number of people who died from causes directly attributable to COVID-19 was $177528 \pm 2790$ and $183821 \pm 1560$ respectively.
The discrepancy between our estimation and those of DPC and ISS could be due to: 
\begin{itemize}
\item not all of the deaths associated with these excesses can be solely attributed to COVID-19. Various concurrent factors contribute to these excess deaths, including increased pressure on the Italian health system. Particularly in the early stages of the pandemic or during certain periods such as the spring of 2021, the healthcare system may have faced challenges in safely treating patients with diseases other than COVID-19. This situation could have resulted in a certain number of individuals not receiving the necessary medical care in hospitals and emergency rooms, leading to additional deaths beyond those directly caused by COVID-19.
\item people with pre-existing serious illnesses, such as heart problems or tumors, may contract COVID-19 and subsequently pass away. However, these deaths are not always classified as deaths directly attributed to COVID-19 since COVID-19 may not be the primary cause of death in such cases. This classification aligns with the definition of death due to COVID-19 provided by the World Health Organization (WHO) in the introduction.
If the same criteria are used to identify a COVID death, then the number of deaths in the ISTAT and ISS data is similar \cite{Cannone};
\item not all the deaths due to COVID-19 have been identified as such. 
\end{itemize}
The temporal correlation between the ISTAT data and the DPC and ISS ones, favours the last point as the dominant as suggested by \cite{Paglino}.

In general, we have no elements to discern this difference and an exhaustive medical discussion is beyond the scope of this article.

We moreover note a large death excess in the 2021 and 2022 summers, probably due to several contiguous heat waves in association with high temperatures.

\subsection{Analysis of the data provided by DPC and ISS}
 \label{par:dpc_iss} 
\begin{figure}[htbp]
    \centering
    \includegraphics[width=.75\textwidth]{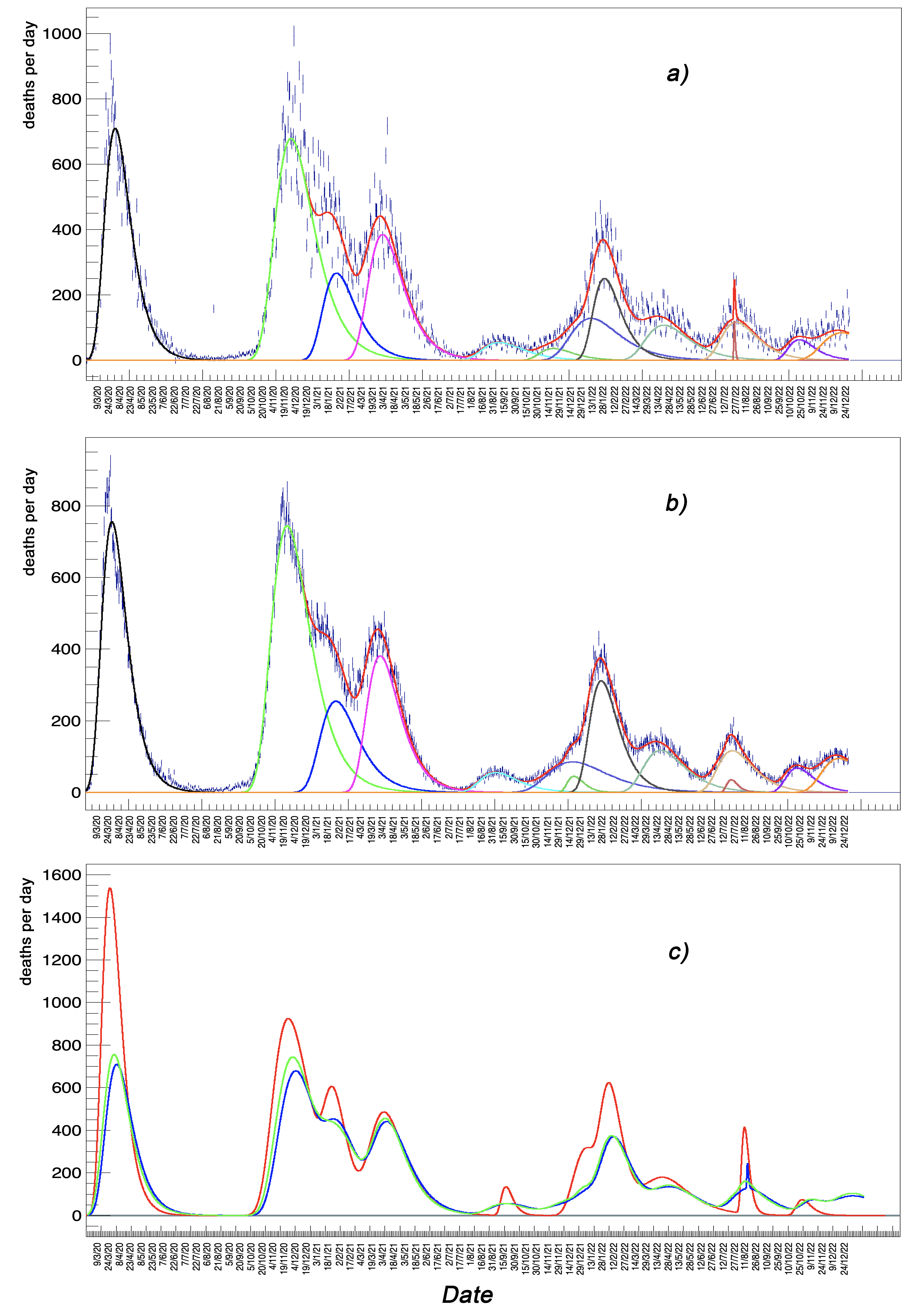}
    \caption {Distribution of deaths due exclusively to the COVID-19 pandemic from February $24^{\mathrm{th}}$ 2020 to December $31^{\mathrm{st}}$ 2022 collected by the DPC ($a$) and ISS ($b$) respectively. In both plots, each COVID-19 wave contribution to the total is reported with different colors, while the overall fit function is plotted in red. In (c) we plot a superimposition of the fit functions relative only to the COVID-19 cause of death, obtained by the interpolation of the ISTAT (red), DPC (blue) and ISS (green) data respectively. The data provided by ISTAT are baseline subtracted.}
    \label{fig:covid_comp}
\end{figure}

The data reported by the DPC and ISS only refer to the deaths directly attributed to COVID-19 using officially defined anamnestic criteria and for this reason have been fitted only with the Gompertz functions relative to the COVID-19 pandemic waves. 
Fig.~\ref{fig:covid_comp} reports the distribution of deaths provided by DPC (a) and ISS (b), relative to the period February $24^{\mathrm{th}}$ 2020 to December $31^{\mathrm{th}}$ 2022) with both the individual Gompertz components and the final overall function, superimposed in red. The statistical fluctuations are very large (especially in the data sample provided by DPC), well above the expected Poissonian ones. Due to the practical impossibility of precisely evaluating the systematic uncertainty associated with the daily number of deaths,  we report only the statistical one. The fit of these distributions requires 13 Gompertz functions, reported in Tab.~\ref{tab:covpar}. 
The fit curves of the DPC and ISS data are almost overlapped and their output parameters are compatible with each other. The number of total deaths results in $177528 \pm 2790$ and  $183821 \pm 1560 $ respectively for DPC and ISS: they are compatible with the ones obtained just summing the number of deaths day per day (without the fit) that is 184918 and 183929 for DPC and ISS respectively.

In Fig.~\ref{fig:covid_comp}c) are reported the overall fit functions of the COVID-19 waves for the ISTAT (red), DPC (blue), and ISS (green) data respectively. In general, for each wave, the number of deaths from DPC and ISS data is lower with respect to the one from ISTAT as already explained in Paragraph~\ref{par:covid_waves} and reported in Table~\ref{tab:covpar}.

\subsection{The quality of the fit}
 \label{par:fit_quality} 
The quality of the fit and the absence of biases can be quantified by the $\chi^{2}/{n_{\mathrm{DOF}}}$ value of the interpolation and by the \emph{pull}, $p_i$, distributions respectively, defined as:
\begin{equation}
    p_i = \frac{d_i - F(t_i)}{\epsilon_i}
\end{equation}
where  $d_{i}$ is the number of death counts in each day $i$, $\epsilon_i$ is its statistical fluctuation ($=\sqrt{d_{i}}$ assuming to follow a Poisson distribution), and $F(t_i)$ is the value of the fit function in that day.
The $\chi^{2}/{n_{\mathrm{DOF}}}$ of the fit of the distribution provided by the ISTAT data (Fig.~\ref{fig:totalfit}) turns out to be $11307.0/4195 = 2.695$, 
indicating a very high quality of the fit considering the large number of degrees of freedom. The value of $\chi^{2}/{n_{\mathrm{DOF}}}$ is significantly different from the expected unity, indicating a non-Poissonian nature of the daily counting of deaths. 

\begin{figure}[htbp]
    \centering
     \includegraphics[width=0.70\textwidth]{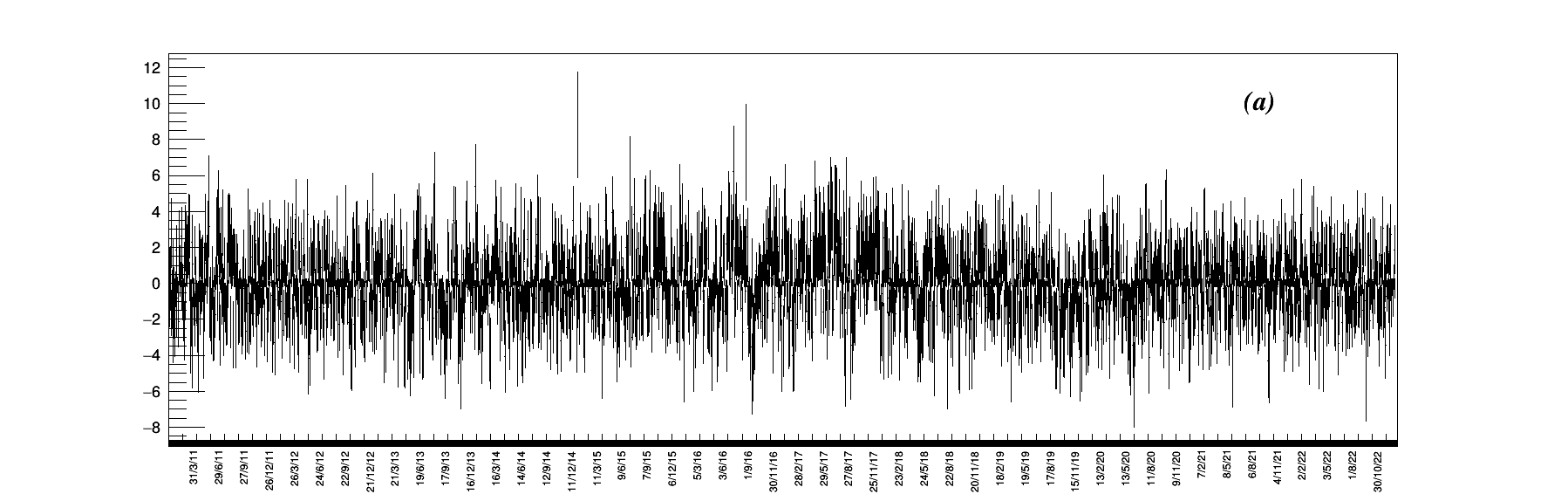}
    \includegraphics[width=0.25\textwidth]{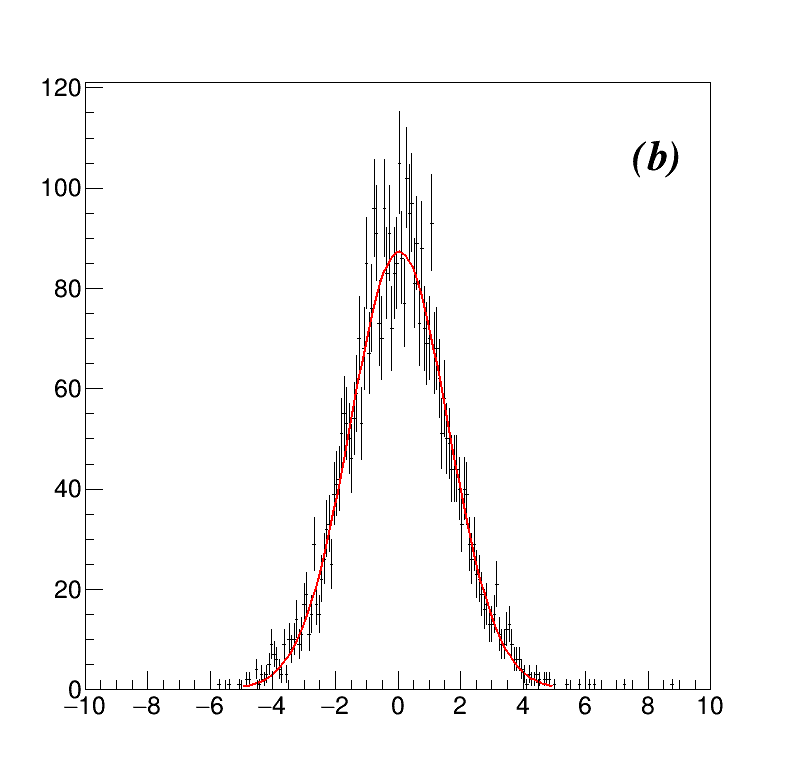}
    \caption{Pull distribution of each day (a) and integrated over the whole period (b).}
    \label{fig:pulls}
\end{figure}

The distributions of the \emph{pull} for each day and integrated over the whole period are reported in Fig.~\ref{fig:pulls}a,b) respectively.
The pull distribution (Fig.~\ref{fig:pulls}a) does not show any systematic deviation from zero to indicate that the fit function is a good approximation of the data for each day. Integrated over the whole period (Fig.~\ref{fig:pulls}b), the mean value and the standard deviation of the pull distribution turn out to be $0.04\pm0.02$ and $1.56\pm0.02$ respectively. The mean value is compatible with zero, while the standard deviation larger than one confirms the underestimation of the uncertainty of the number of deaths each day: the Poissonian uncertainty associated with the number of deaths correctly takes into account the independence of the counts but does not consider any effects that can correlate the deaths with each other, such as strong flu that can saturate hospitalizations and consequently increase the number of deaths. This deviation from unity, of about $60\%$, gives an approximate amount of the increase that could be applied to the data errors to make them compatible with Poissonian values. The quality of the fits of the data from ISTAT, together with the one from DPC and ISS data, is summarized in Table~\ref{tab:qfit}.
 \begin{table}[htbp]
 \centering 
 \begin{tabular}{|| c | c | c | c | c | c ||} 
  \hline
  \emph{DATA}  & \emph{points} & \emph{NDF}  & \emph{$\chi^{2}/{n_{\mathrm{DOF}}}$} & \emph{pull: $\mu$} &\emph{pull: $\sigma $} \\ [0.5ex] 
  \hline\hline 
  ISTAT   &   4381   &   4195   &  2.695  &  $0.04 \pm 0.02 $   &   $1.56 \pm 0.02 $  \\
  DPC     &   1041   &   1002   &  10.90  &  $1.2 \pm 0.1 $     &   $2.4 \pm 0.1 $    \\
  ISS     &   1041   &   1002   &  3.13   &  $0.35 \pm 0.05 $   &   $1.63 \pm 0.05 $  \\
  \hline
 \end{tabular}
 \caption{Main parameters to quantify the quality of the fit for ISTAT, DPC, and ISS data respectively. The \emph{$\chi^{2}/{n_{\mathrm{DOF}}}$} and the pull of the DPC and ISS (to a lesser extent) show values above the ones expected by the only Poissonian fluctuations and possible explanations are given in~\ref{par:dpc_iss})}
 \label{tab:qfit}
 \end{table}

\subsection{Evaluation of the systematic errors}
\label{par:systematic} 

The sources of systematic errors in extracting the excesses from a least-squares fit technique depend on both the data acquisition method and the subsequent analysis. The ISTAT data do not exhibit systematic effects since they provide the exact date of death and encompass the entirety of deaths at the national level for all the considered years. Therefore, the ISTAT data can be considered reliable and free from systematic errors in this context.

The choice of the fit function is indeed a significant factor contributing to systematic errors in the analysis method. In addition to the fit function expressed in $F_{ref}(t)$ (eq. ~\ref{fullFitFunction}), we have considered two additional functions. 

In the second fit approach, each excess of death has been modeled by a Gaussian function $g_j(t)$. The overall fit function is expressed in the following form:
\begin{equation}
  \label{fullFitFunction2}
  F_2(t)=baseline(t)+
  {\sum_{j=1}^{n}g_{j}(t)}
\end{equation}

where the $baseline$ is the same as ~\ref{fullFitFunction}.
The choice of the Gaussian function is justified because it represents a symmetrical distribution of a random variable. 

In our third fit attempt, we have included an additional hypothesis to the definition of death excess; all the excesses arisen by causes such as an epidemic or flus have been parametrized by a Gompertz function, while the other death excesses such as heat or cold waves have been parametrized by a Gaussian.
Finally, causes of death such as car accidents, work-related or other similar sources are completely random on each given day of the year and are therefore represented by a constant term. 

By adopting the criteria highlighted above, we ended up considering 22 peaks originated by epidemic origin, interpolated with Gompertz functions, and 38 peaks with a symmetric structure due to unrelated, random, events, thus interpolated with Gaussian functions.
The final fit function has therefore been expressed in the following form:
\begin{equation}
  \label{fullFitFunction3}
  F_3(t)=baseline(t)+
  {\sum_{j=1}^{l}\dot{G}_{j}(t)}+
  {\sum_{i=1}^{m}g_{i}(t)}
\end{equation}

where the meaning of the symbols is the same as equations ~\ref{fullFitFunction},~\ref{baseline}, ~\ref{eq:gomp}. 
In all three fit variants described above to assess the systematic effects, the number of free parameters and consequently the degrees of freedom of the fits are all the same. 
Furthermore, due to the importance of the baseline in the analysis, we have included in the systematic error evaluation the contribution resulting from variations in the baseline parameters, in particular, the distribution of deaths was calculated as follows:
\begin{itemize}
\item by freezing the baseline parameters and changing only those related to the Gompertz functions;
\item by freezing the baseline parameters obtained by fitting until January $31^{\mathrm{st}}$ 2020 just before the outbreak of the pandemic. 

\item replacing the expression of the baseline as in the formula:
\begin{equation}
  \label{baselinep}
  baseline^{'}(t)=const + slope\cdot t + A_0 \sin\left(\frac{2\pi t}{T}+\varphi\right)\,
\end{equation}
with a constant sine amplitude.
\end{itemize}

For each excess, the systematic uncertainty associated with the different fit functions has been estimated as the square root of the quadratic distance between the number of excess deaths determined by each fit and the simple average of the three methods, by the formula:
\begin{equation}
  \label{eq:fitvariant}
  \sigma_{fit}=\sqrt{\frac{{\sum_{i=1}^{N} x_i^2 -N{<x>}^2  }}{N-1}}
\end{equation}
where $N$ is the number of fit variants (6 in this case) and, for each excess, $x_i$ is the number of deaths, $<x>$ is the average of the six variants (for more details see~\cite{fitvarian}).
The systematic uncertainty has been evaluated only for the whole sample of data from ISTAT and it has been reported in Tables~\ref{tab:wint_sum}, ~\ref{tab:ave_meth} and ~\ref{tab:covpar}. For each excess, the systematic uncertainty is at most 10-20$\%$. 

\section{Mortality by gender and class age}
\begin{figure}[htbp]
    \centering
    \includegraphics[width=0.99\textwidth]{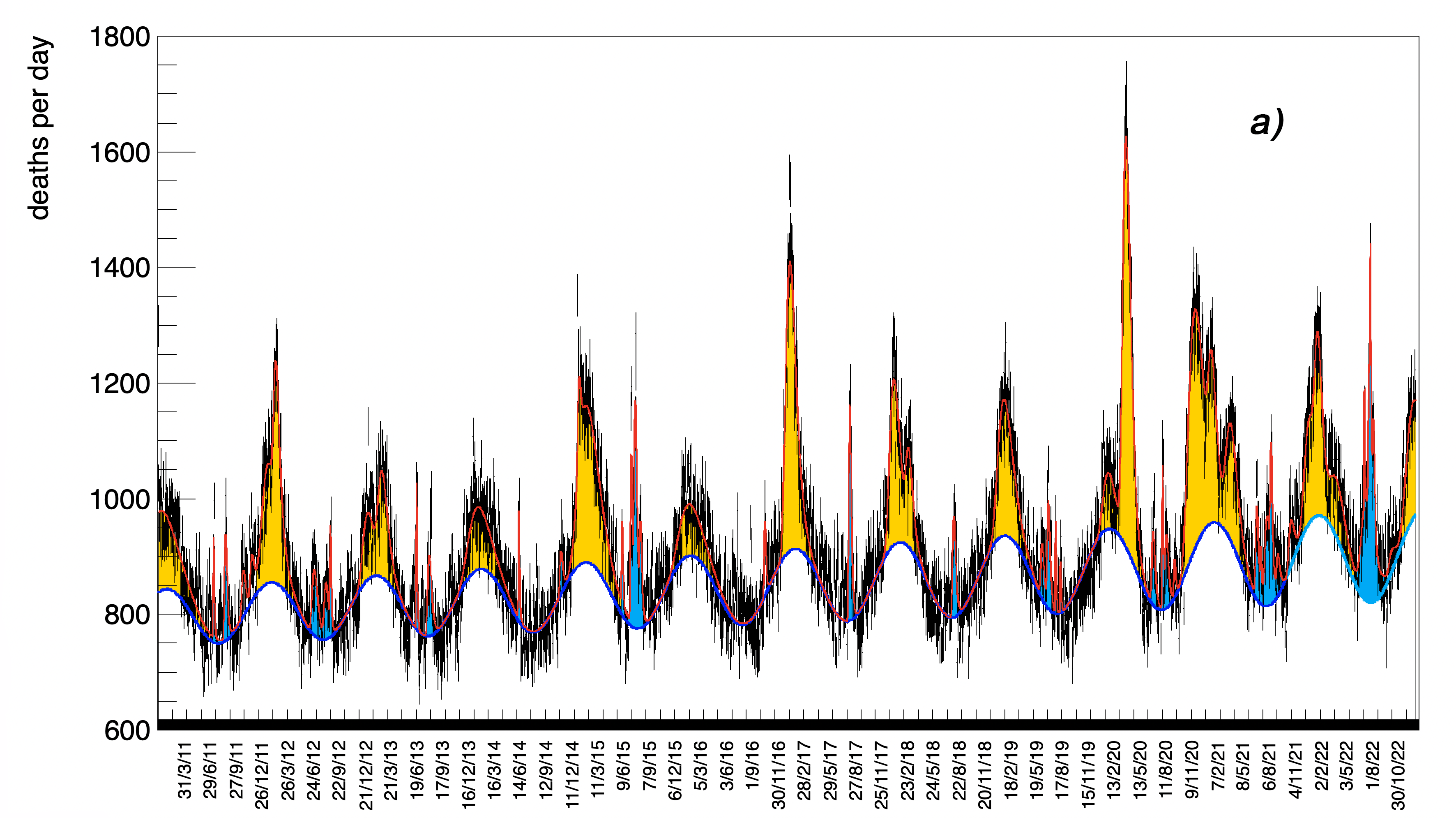}
    \includegraphics[width=0.99\textwidth]{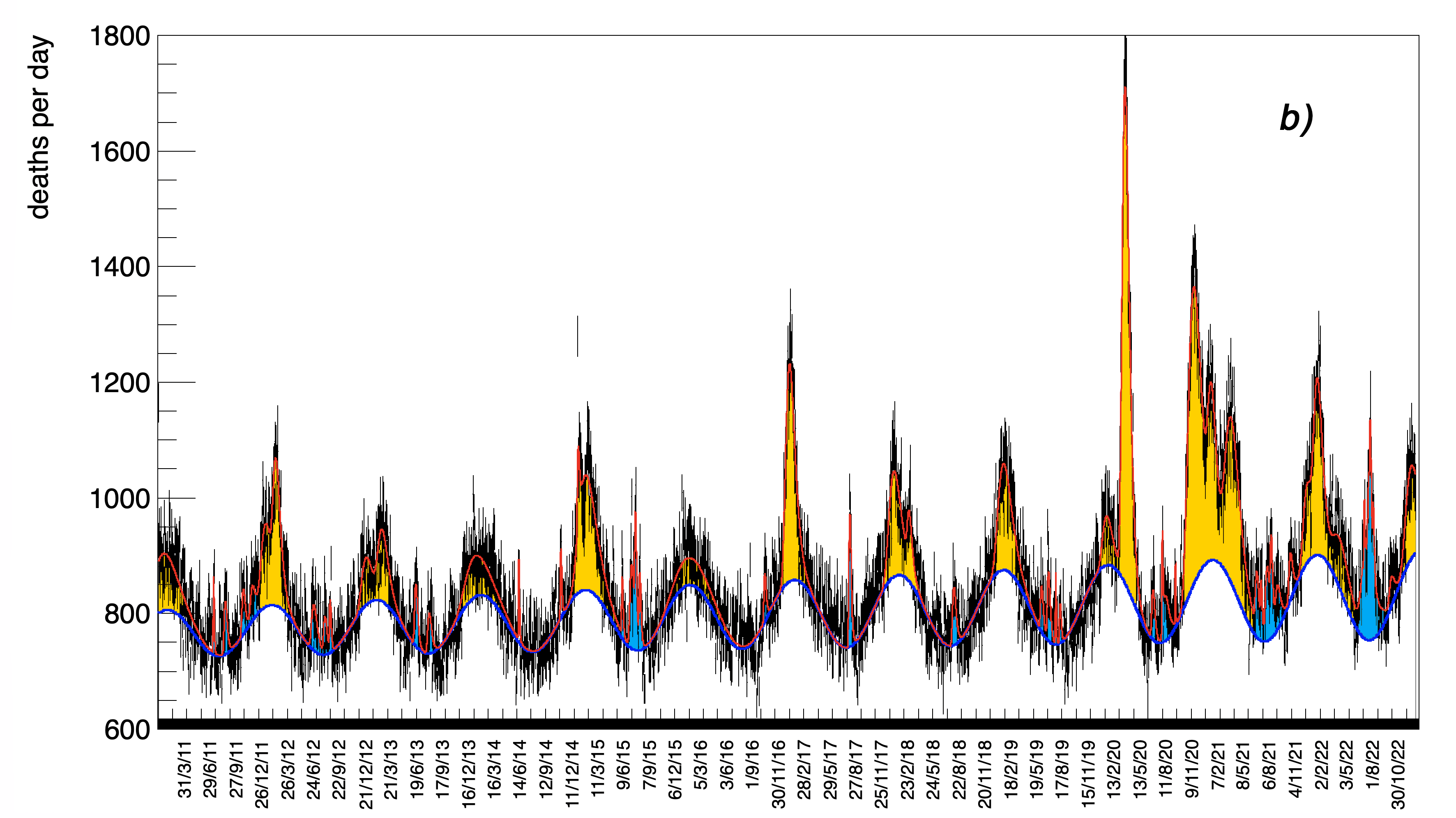}
    \caption{Distribution of deaths collected by ISTAT in Italy between January $1^{\mathrm{st}}$ 2011 to December $31^{\mathrm{th}}$ 2022 for women a) and men b). These data include all classes of ages. As already mentioned, the yellow areas are the mortality excesses evaluated by the fit function above the baseline for the winter periods while the cyan ones are for the summer periods.}
    \label{fig:wom_men}
\end{figure}

To ascertain whether COVID-19 affects the most a particular gender or age class, the whole data from ISTAT has been disentangled into different sub-samples. In Fig.~\ref{fig:wom_men} the distribution of deaths is shown for women and men respectively, fitted with the same function as eq.~\ref{fullFitFunction}.

The characteristics of the baseline resulting from fitting these disentangled data samples are reported in Table~\ref{table:base_wom_men}

 \begin{table}[htbp]
 \centering
 \begin{tabular}{|| c |  c | c | c | c | c | c ||} 
  \hline
   \emph{$gender$} &\emph{$const$} & \emph{$slope$} &\emph{$A_0$} &\emph{$A_1$} & \emph{$T$} (days) & \emph{$\varphi$} (rad)\\ [0.5ex] 
  \hline\hline
 women & 790.0 $\pm$ 3.3 & 0.026 $\pm$ 0.001 & 47.6 $\pm$ 4.1 & 0.007 $\pm$ 0.002 & 365.7 $\pm$ 0.4 & -5.13 $\pm$ 0.04 \\ 
 men   & 765.1 $\pm$ 2.4 & 0.015 $\pm$ 0.001 & 40.6 $\pm$ 3.0 & 0.008 $\pm$ 0.001 & 364.4 $\pm$ 0.3 & -5.24 $\pm$ 0.04 \\ 
 age $\geq$ 60 & 1425.2 $\pm$ 3.4 & 0.044 $\pm$ 0.001 & 90.8 $\pm$ 4.9 &  0.015 $\pm$ 0.002 & 364.8 $\pm$ 0.2 & -5.23 $\pm$ 0.02 \\ 
 \hline
 \end{tabular}
 \caption{Characteristics of the baseline for different genders (all age classes) and the age class over or equal to 60 years old. }
 \label{table:base_wom_men}
 \end{table}

The difference of about 3\% in average daily death (parameter $const$ of the function in Table~\ref{table:base_wom_men}) between women and men, reflects the larger fraction of the female population (in 2021 was 51,3\%), with respect to the male one (48,7\%). The total number of deaths due to the COVID-19 pandemic waves is $100028 \pm 1799$ and $114538 \pm 1441$ for women and men respectively, corresponding to a percentage of $47 \pm 1\%$ (female) and $53 \pm 1\%$ (male). The lower percentage of dead women compared to men due to the COVID-19 pandemic is in opposition to the higher percentage of the female population in the most affected age class by COVID-19 disease ($\geq60$ years old): the lower female mortality could be due to various factors such as
any differences in the comorbidity.
By defining the average gender mortality as the ratio of the number of deaths and the relative gender population (we took the one relative to January $1^{\mathrm{st}}$ 2023) in the most affected age class ($\geq60$ years old), we obtain $1.00 \pm 0.02\%$ and $1.39 \pm 0.02\%$ for female and male respectively, confirming evidence of a higher excess of deaths in the male gender due to the COVID-19 pandemic. 
All the excesses of deaths for each pandemic wave are reported in Table~\ref{tab:covid_wom_men}. 

 \begin{table}[htbp]
 \centering 
 \begin{tabular}{|| c | c | c | c | c | c ||} 
  \hline
  \emph{COVID-19 wave}  & \emph{function}   & \emph{deaths Women}  &\emph{deaths men} &\emph{deaths age $\geq$ 60} &\emph{deaths age 30-59}  \\ [0.5ex] 
  \hline\hline 
  March 2020   &   $\dot{G}_{37}$   &   $25675 \pm 406  $   &  $27689 \pm 332 $   &   $52259 \pm 477 $  &   $1356 \pm 697 $\\
  November 2020   &   $\dot{G}_{42}$   &   $29536 \pm 702  $   &  $31677 \pm 500 $   &   $58927 \pm 691 $  &   $1928 \pm 1015 $\\
  January 2021   &   $\dot{G}_{43}$   &   $6969 \pm 472  $   &  $8902 \pm 440 $   &   $15221 \pm 586 $  &   $1127 \pm 832 $\\
  March 2021   &   $\dot{G}_{44}$   &   $9651 \pm 531  $   &  $18489 \pm 564 $   &   $25910 \pm 732 $  &   $2356 \pm 1065 $\\
  August 2021   &   $\dot{G}_{49}$   &   $1004 \pm 127  $   &  $1679 \pm 241 $   &   $2367 \pm 336 $  &   $333 \pm 406 $\\
  December 2021   &   $\dot{G}_{51}$   &   $10496 \pm 957  $   &  $6461 \pm 643 $   &   $16865 \pm 1700 $  &   $704 \pm 2334 $\\
  January 2022   &   $\dot{G}_{52}$   &   $7276 \pm 754  $   &  $10683 \pm 574 $   &   $16392 \pm 1393 $  &   $1063 \pm 1964 $\\
  March 2022   &   $\dot{G}_{53}$   &   $5695 \pm 657  $   &  $6122 \pm 576 $   &   $12112 \pm 743 $  &   $582 \pm 1177 $\\
  July 2022   &   $\dot{G}_{57}$   &   $2790 \pm 251  $   &  $1850 \pm 178 $   &   $4557 \pm 628 $  &   $731 \pm 1213 $\\
  October 2022   &   $\dot{G}_{59}$   &   $936 \pm 286  $   &  $987 \pm 198 $   &   $1890 \pm 284 $  &   $0 \pm 411 $\\
 \hline
 All waves  &      &   $100028 \pm 1799  $   &  $114538 \pm 1441 $   &   $206501 \pm 2748 $  &   $10176 \pm 3980 $ \\
  \hline
 \end{tabular}
 \caption{Number of deaths for each COVID-19 pandemic wave for women, men, and the age class older than 60 years old. The last column contains an estimation of the number of deaths in the middle-class age obtained by subtracting the old-class age from the total one and supposing that the COVID-19 pandemic did not cause any deaths in the young-class age.}
 \label{tab:covid_wom_men}
 \end{table}

The sum of the number of deaths of the two genders is compatible with that of the total sample (see the fourth column of Table~\ref{tab:covpar}) giving further confirmation of the goodness and self-consistency of the fit results. 
The quality of the fit is good for both gender, the $\chi^{2}/{n_{\mathrm{DOF}}}$ is 2.00 and 1.77 for women and men respectively; the pull distributions present a mean value of $ 0.03 \pm 0.02 $ and $ 0.04 \pm 0.02 $ and a standard deviation of $ 1.33 \pm 0.02 $ and $ 1.28 \pm 0.02 $ for female and male gender respectively.

It is possible to compare these results with the ones provided by the ISS.
Without using the fit results, but by just summing the number of deaths for each given day, ISS has registered 75757 and 95405 deaths for female and male genders respectively due to the COVID-19 pandemic period. 
With respect to the total number of deaths, the female percentage is 44.2\% and consequently, the male is 55.8\%.
As we discussed above, the absolute number of deaths is underestimated with respect to the one from ISTAT, but the death percentage for each gender is completely in agreement.

\begin{figure}[htbp]
    \centering
     \includegraphics[width=0.85\textwidth]{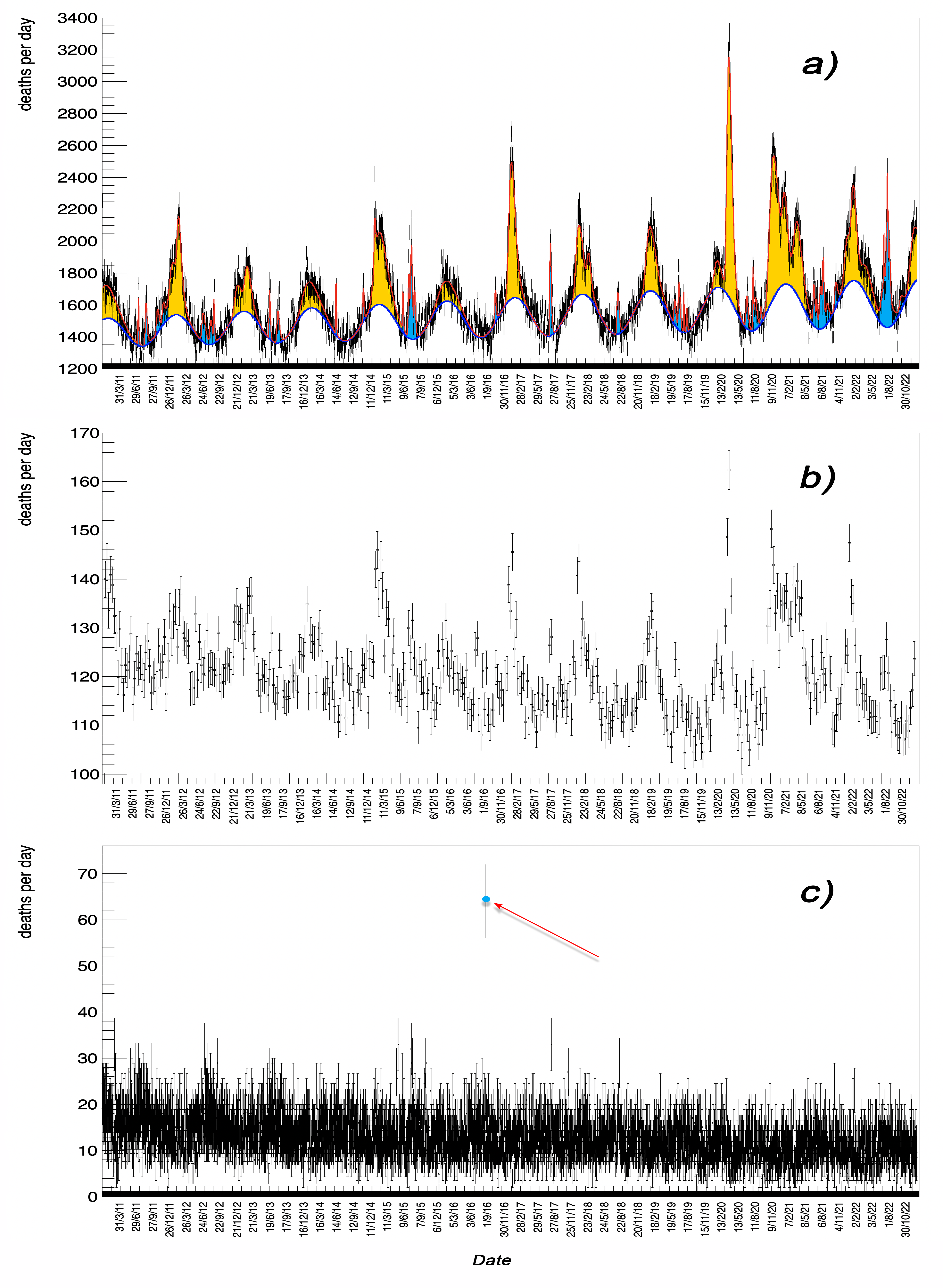}
    \caption{Distribution of deaths collected by ISTAT in Italy between January $1^{\mathrm{st}}$ 2011 to December $31^{\mathrm{th}}$ 2022 for three different age classes: from the top to bottom, older than 60 (old age class), from 30 to 59 (middle age class) and less than 30 (young age class) years old. The middle-age class plot has been averaged into a time interval of 10 days to provide a clearer view of the individual excess peaks. Red arrow in panel c) as in Fig.~\ref{fig:totalnofit} panel b).}
    \label{fig:age}
\end{figure}

We have also disentangled the full sample of data into 3 contiguous age classes; from 0 to 29 (young age class), from 30 to 59 (middle age class), and equal to or more than 60 years (old age class): their percentage in the Italian population of 2021 is 27,8\%, 42,0\% and 30.2\% respectively.
The distribution of deaths is shown in Fig.~\ref{fig:age}.

The typical periodic pattern visible in the death distribution of the old age class (Fig.~\ref{fig:age} a) is essentially due to people who are generally weaker and therefore more sensitive to seasonal weather variations and flu epidemics. 
The old-age class has been fitted with the same function as eq.~\ref{fullFitFunction}. The quality of the fit is good: the $\chi^{2}/{n_{\mathrm{DOF}}}$ is 2.670 and the pulls present a mean value of $ 0.03 \pm 0.02$ with a standard deviation of $1.55 \pm 0.02$. The average number of daily death is $1425.2 \pm 3.5$ which corresponds to 91.2\% of the total; all the details of the baseline are shown in Table~\ref{table:base_wom_men}.
The effect of the COVID-19 pandemic stands out with a total number of deaths of $206501 \pm 2748$ (95.3 \% of the total), corresponding to an average mortality of $1.13 \pm 0.02\%$ in this age class. 
The percentage of deaths in this age class is in complete agreement with the one evaluated from the ISS data which is 94.9\%.

In the middle-aged classes, the periodic daily death distribution is barely visible as are any excess peaks, a fact which prevents fitting the distribution with the eq.~\ref{fullFitFunction}. The rather large statistical fluctuations hamper the possibility of significantly identifying any periodical structure or excess of death: for this reason, in Fig.~\ref{fig:age} b) the number of daily deaths has been averaged in a time interval of 10 days, allowing better visual identification of excess of death peaks.
Due to the impossibility of obtaining a good quality fit of the middle-class age distribution to estimate the COVID-19 effect on mortality, the best approximation is by subtracting from the total number of deaths the ones from the old age class (both fits are very precise) and then to assume that there have been no deaths due to COVID-19 pandemic among the young age class (see Fig.~\ref{fig:age} c). With this method, the total number of deaths in the 10 COVID-19 pandemic waves is $10176 \pm 3980 $ and the details on every single wave are given in Table~\ref{tab:covid_wom_men}.
The number of deaths for each COVID-19 pandemic wave of the old and middle age classes is reported in Table~\ref{tab:covid_wom_men}. 

The daily death of the young age class presents a rather flat distribution, without any significant deviation from it.  
The effect of COVID-19 on the young age class is practically zero (no excess in Fig.~\ref{fig:age} c).
To get a rough estimate of the average number of daily deaths and the eventual slope of the overall trend, we have fitted both the last two age class distributions with a simple straight line. For the young age class we have obtained $const = 14.3 \pm 0.1$ and $slope = (-1.30 \pm 0.04) \cdot 10^{-3}$ (same notation as formula~\ref{baseline}), with a $\chi^{2}/{n_{\mathrm{DOF}}}$ of 1.26. The middle age class provides $const = 125.3 \pm 0.3$ and $slope = (-2.6 \pm 0.1) \cdot 10^{-3}$, with a $\chi^{2}/{n_{\mathrm{DOF}}}$ of 1.56: the higher value of $\chi^{2}$ reflects the significant deviation of the distribution from a straight line.
In the last 12 years, in the young-age class, the average daily death decreased from 14.3 to 9 only in part due to the demographic decrement of about 10\% for that time period; in the middle-aged class instead, the average daily death decreased by 7\% (from 125 to 116) perfectly in line with the 6\% of demographic decline in this class age. 

\section{Conclusions}

After approximately three years of the COVID-19 pandemic, the need for a reliable, accurate, and statistically robust method to estimate excess deaths has become crucial. To address this need, we have analyzed death data from multiple sources from January $1^{\mathrm{st}}$ 2011 to December $31^{\mathrm{th}}$ 2022. Firstly, we examined data from ISTAT, which includes information on deaths from all causes without specific reference to any particular one, in Italy over the past 12 years. Additionally, we analyzed data provided by the Italian Department for Civil Protection (DPC) and the Italian National Institute of Health (ISS) specifically related to deaths directly attributed to COVID-19 since the onset of the pandemic.

For a correct evaluation of the death excess, it is necessary an accurate determination of the baseline representing the average number of expected deaths each day in absence of any seasonal excess, which is the main focus of our analysis. The baseline shows a periodical behavior well described by a sinusoidal function in phase with the yearly seasonal cycle, plus a linear function. The determination of the excesses of death has been evaluated by a statistical interpolation of the data based on a $\chi^2$ minimization method using a function that is the sum of the baseline and 60 Gompertz functions to represent each excess of deaths. We are then able to correctly estimate the excess of deaths for each COVID-19 pandemic wave, even if these are partially overlapped.
The quality of the fit is satisfactory both in terms of $\chi^2$  and pull distributions.
It has been established that the total number of excess deaths due to the COVID-19 pandemic or related indirect effects is $216677 \pm 2878 (\text{(statistical)} \pm 6843\text{(systematic)}$. 
This figure exceeds those obtained from data provided by the ISS and DPC by approximately 35,000 cases.
We have demonstrated that the excess deaths are strongly correlated in time with the COVID-19 peaks. 

About 50\% of the deaths occurred in the first two COVID-19 waves when the national health system was not yet prepared to deal with an epidemic of this magnitude, the vaccine was not yet ready and probably the mortality rate was higher than the following.

We have divided the Italian population into three age classes (0-29, 30-59, and greater than or equal to 60 years): the younger age class does not present any excess death during the COVID-19 pandemic, while the number of deaths in the other two age classes resulted in about 5\% and 95\% of the total, respectively.
Considering the most affected class age ($\geq 60$ years old), we have confirmed evidence of greater mortality in the male gender. Various excesses of death that occurred in summer (winter) in a very short period due to excesses of heat (cold) were also identified and quantified.

\section{Data Availability Statement}
This study is based on three different sets of data provided by:
\begin{itemize}
\item ISTAT (Italian National Institute of Statistics) available at \cite{ISTAT-Data}

{\it \link{https://www.istat.it/it/archivio/240401}}

Data can be downloaded in excel format in the subcategory "Decessi anni 2011-2024" 

\item DPC (Italian Department for Civil Protection) available at \cite{ProtezioneCivile-Data} 

{\it \link{https://github.com/pcm-dpc/COVID-19/tree/master/dati-andamento-nazionale}}
\item ISS (Italian National Institute of Health) available in an integrated form at \cite{ISS-data} 
 
{\it \link{https://www.epicentro.iss.it/coronavirus/sars-cov-2-sorveglianza/}}

and on a daily basis at \cite{covidstat} 

{\it \link{
https://covid19.infn.it/}} 

through an agreement between the ISS and the National Institute of Nuclear Physics (INFN).
\end{itemize}

\section{Acknowledgements}
The authors from the National Institute for Nuclear Physics (INFN) would like to express their gratitude to their colleagues from the National Institute of Health (ISS) for allowing the use of their data on the COVID-19 epidemic. This analysis represents the outcome of a collaborative agreement between the two research institutes.

\end{document}